\documentclass[10pt,doublecolumn]{IEEEtran}
\usepackage{amsmath,amsthm,amsfonts,amssymb,bm,mathrsfs}
\usepackage{graphicx,subfigure}
\usepackage[cmintegrals]{newtxmath}
\usepackage{amssymb}
\usepackage{cite}
\usepackage{setspace}
\usepackage{caption}
\usepackage{floatrow}
\usepackage{algorithm}
\usepackage{algorithmic}
\usepackage{float}
\ifCLASSINFOpdf
\else
\fi
\begin{document}
	
%
\title{Fundamentals on Base Stations in Cellular Networks: From the Perspective of Algebraic Topology}

\author{Ying Chen, \IEEEauthorblockN{Rongpeng Li, Zhifeng Zhao, and Honggang Zhang}\\
	\thanks{Y. Chen, R. Li, Z. Zhao, and H. Zhang are with College of Information Science and Electronic Engineering, Zhejiang University. Email: \{21631088chen\_ying, lirongpeng, zhaozf, honggangzhang\}@zju.edu.cn}
	\thanks{This work was supported in part by National Key R$ \& $D Program of China (No. 2018YFB0803702), National Natural Science Foundation of China (No. 61701439, 61731002), Zhejiang Key Research and Development Plan (No. 2018C03056), the National Postdoctoral Program for Innovative Talents of China (No. BX201600133), and the Project funded by China Postdoctoral Science Foundation (No. 2017M610369).}}


%


\maketitle

\begin{abstract}
 In recent decades, the deployments of cellular networks have been going through an unprecedented expansion. In this regard, it is beneficial to acquire profound knowledge of cellular networks from the view of topology so that prominent network performances can be achieved by means of appropriate placements of base stations (BSs). In our researches, practical location data of BSs in eight representative cities are processed with classical algebraic geometric instruments, including $ \alpha $-Shapes, Betti numbers, and Euler characteristics. At first, the fractal nature is revealed in the BS topology from both perspectives of the Betti numbers and the Hurst coefficients. Furthermore, log-normal distribution is affirmed to provide the optimal fitness to the Euler characteristics of real BS deployments.
\end{abstract}


\section{Introduction}
Prompted by the significance of base station (BS) deployment issues, substantial researchers have been working on the relevant subjects over decades \cite{Kibi2016Modelling,Zhou2016Large,Li2016On,Zhao2017Temporal,Kwon2015Random}.

However, the majority of related documents studied BS deployments only by means of analyzing BS density distributions. In our works, unprecedentedly, several principal concepts in algebraic geometry field, i.e., $ \alpha $-Shapes, Betti numbers, and Euler characteristics \cite{Weygaert2011Alpha}, are merged into the analyses of the BS topology of eight representative cities around the world. In essential, BSs can be abstracted into a discrete point set in the 2-dimensional plane, and connections between BSs can be specifically established according to the construction of $ \alpha $-Shapes, so $ \alpha $-Shapes are capable of reflecting topological features of BS deployments. Moreover, the topological information can be extracted from the Betti numbers and the Euler characteristics as well, because the corresponding $ \alpha $-Shapes can be expressed in terms of these two features. Due to the space limitation, interesting readers could refer to \cite{Chen2018Topology} to find the details of these three topological notions and their relationships.

To be specific, this letter is devoted to searching for essential topological features in BS deployments, which transcend geographical, historic, or culture distinctions, and these properties stand a good chance to provide valuable guidance for designers of cellular networks. For this purpose, this letter offers two novel contributions as listed below:
\begin{itemize}
	\item Firstly, the fractal nature is uncovered in BS configurations based on the Betti numbers and the Hurst coefficients;
	\item Secondly, log-normal distribution is confirmed to match the Euler characteristics of the real BS location data with the best fitting among all candidate distributions.
\end{itemize}

The organization of this letter is arranged as follows: The actual BS location data are briefly introduced in Section II. The fractal phenomenon in BS deployments for either Asian or European cities is presented in Section III. The log-normal distribution is affirmed to offer the best fitting for the Euler characteristics of all the eight cities in Section IV. Lastly, conclusions are summarized in Section V.

\section{Dataset Description}
A great deal of real data downloaded from \emph{OpenCellID} community (https://community.opencellid.org/) \cite{Ulm2015Characterization}, a platform to provide BS's latitude and longitude data all over the world, are analyzed to guarantee the validity of results. Substantial BS location data of eight cities, including four representative cities in Asia (i.e., Seoul, Tokyo, Beijing, and Mumbai) and Europe (i.e., Warsaw, London, Munich, and Paris) respectively, have been collected from this platform. The basic information of these eight cities is provided in Table I.

\section{Fractal Nature in Cellular Networks Topology}
As a vital property of complex networks, the fractal nature has been revealed in a plenty of wireless networking scenarios \cite{Yuan2017The,Ge2016Wireless,Hao2017Wireless,Strogatz2005Complex,Song2005Self}. In terms of the Betti numbers and the Hurst coefficients, the fractal feature in the BS topology is verified in this section. 

\subsection{Fractal Features Based on the Betti Curves}
The random and fractal point distributions can be distinguished by their Betti curves as depicted in Fig. 1 (for details see \cite{Pranav2017The}).

\begin{figure}[htbp]
	\centering
	\subfigure[Points deployment diagrams: (left) random; (right) fractal.]{
		\makebox[4cm][c]{
			\includegraphics[scale=0.13]{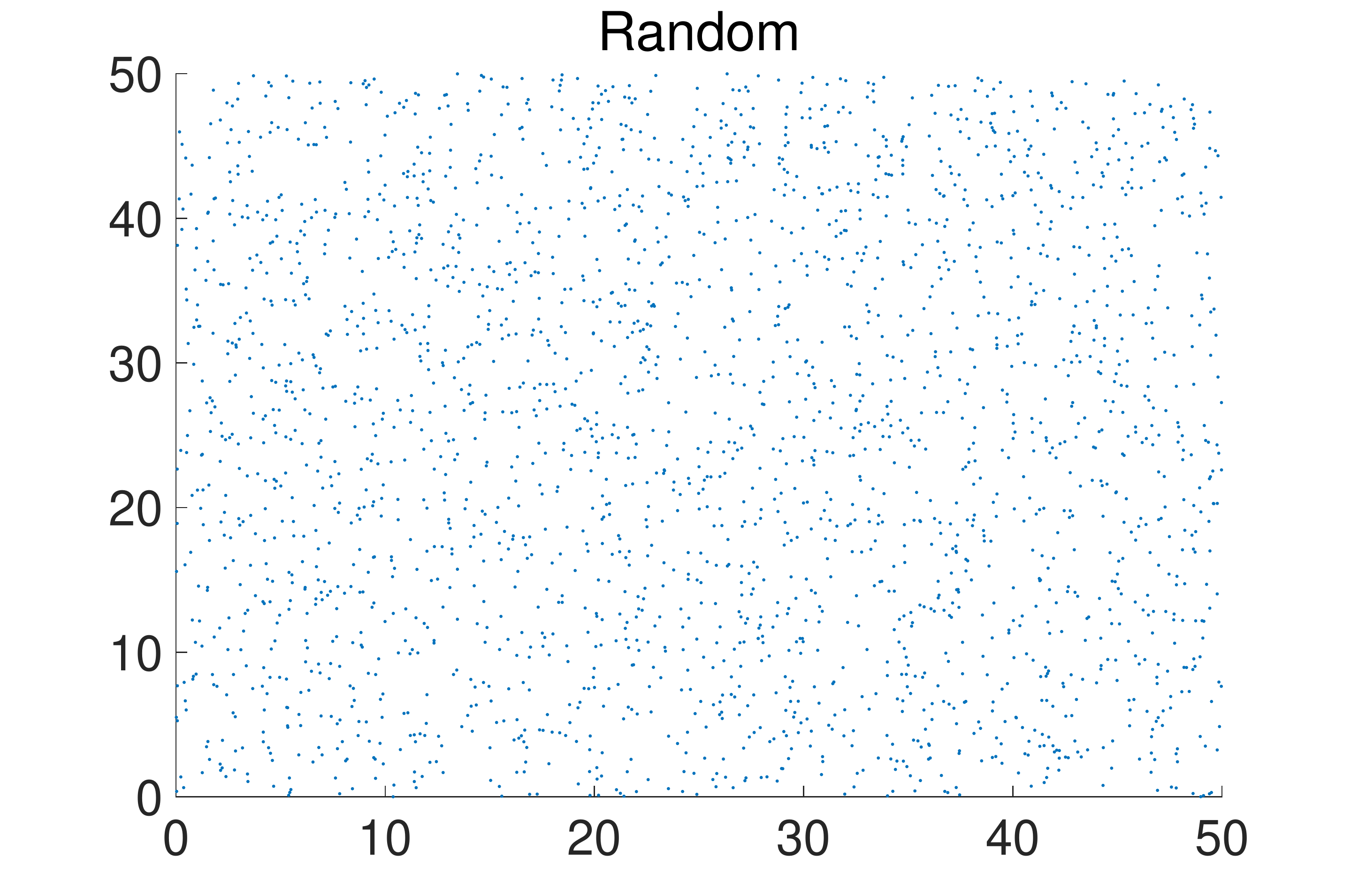}
		}
		\makebox[4cm][c]{
			\includegraphics[scale=0.13]{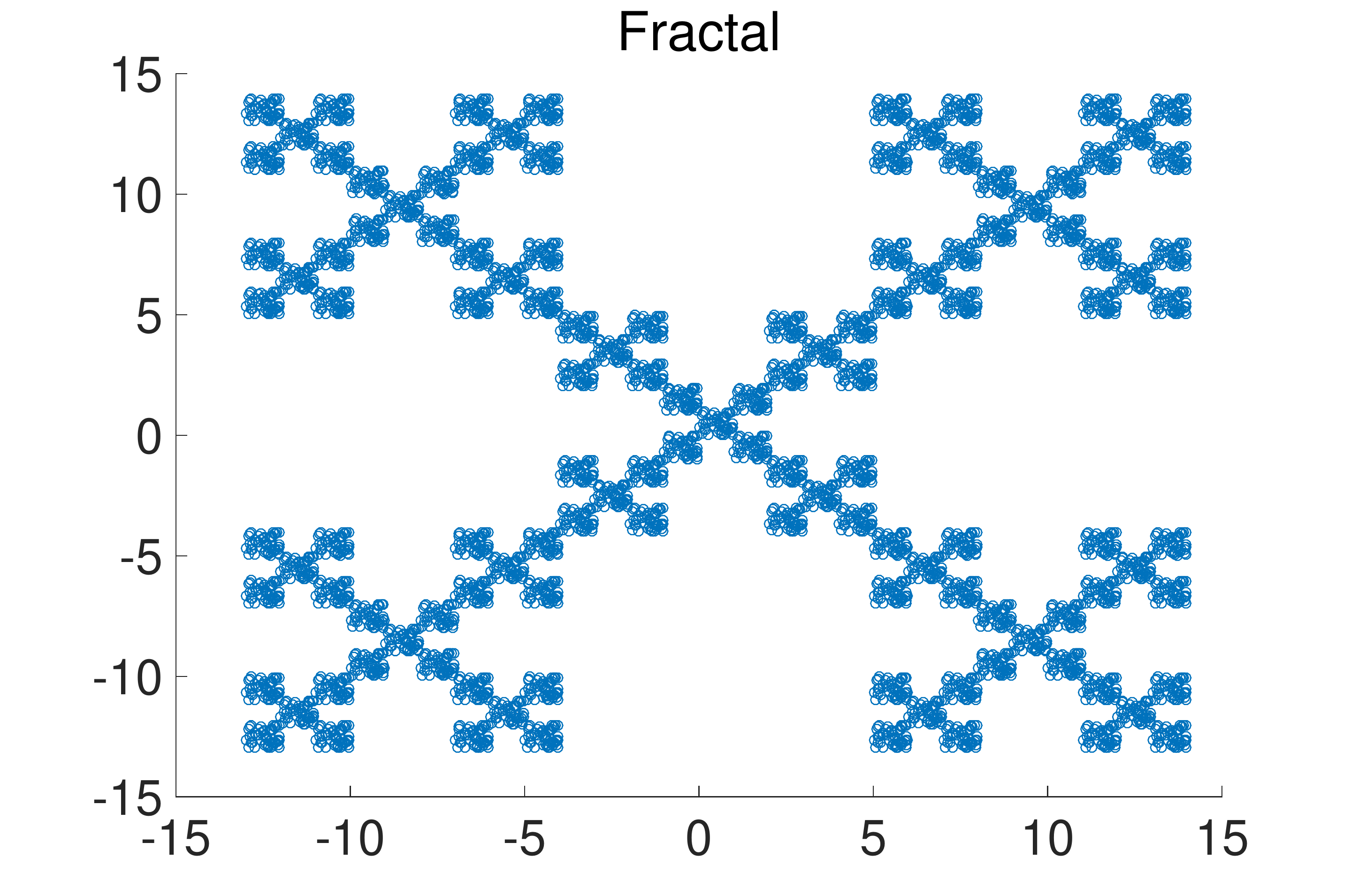}
		}
	}

	\subfigure[Betti curves for random: (left) $ \beta_{0} $; (right) $ \beta_{1} $.]{
		\makebox[4cm][c]{
			\includegraphics[scale=0.13]{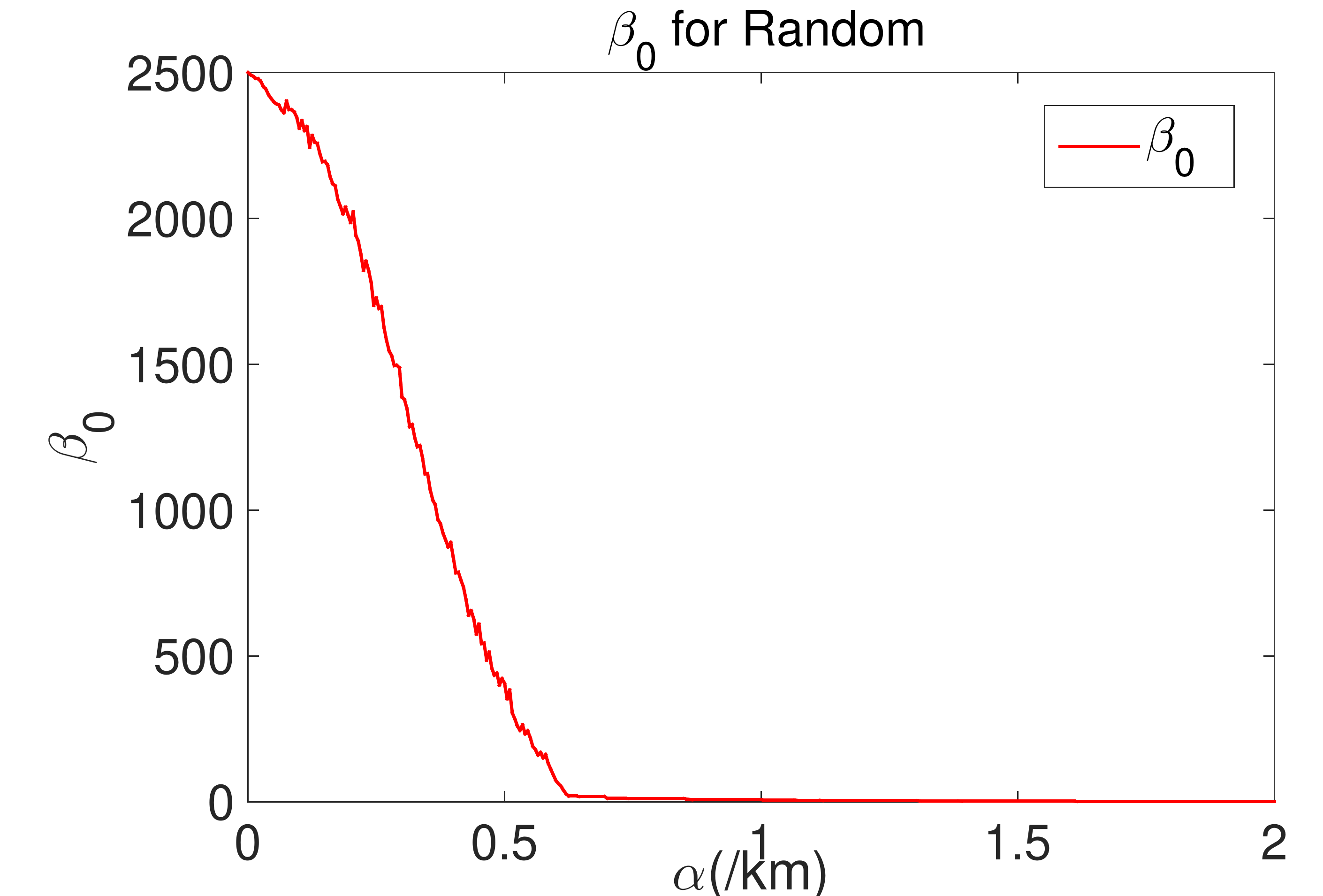}
		}
		\makebox[4cm][c]{
			\includegraphics[scale=0.13]{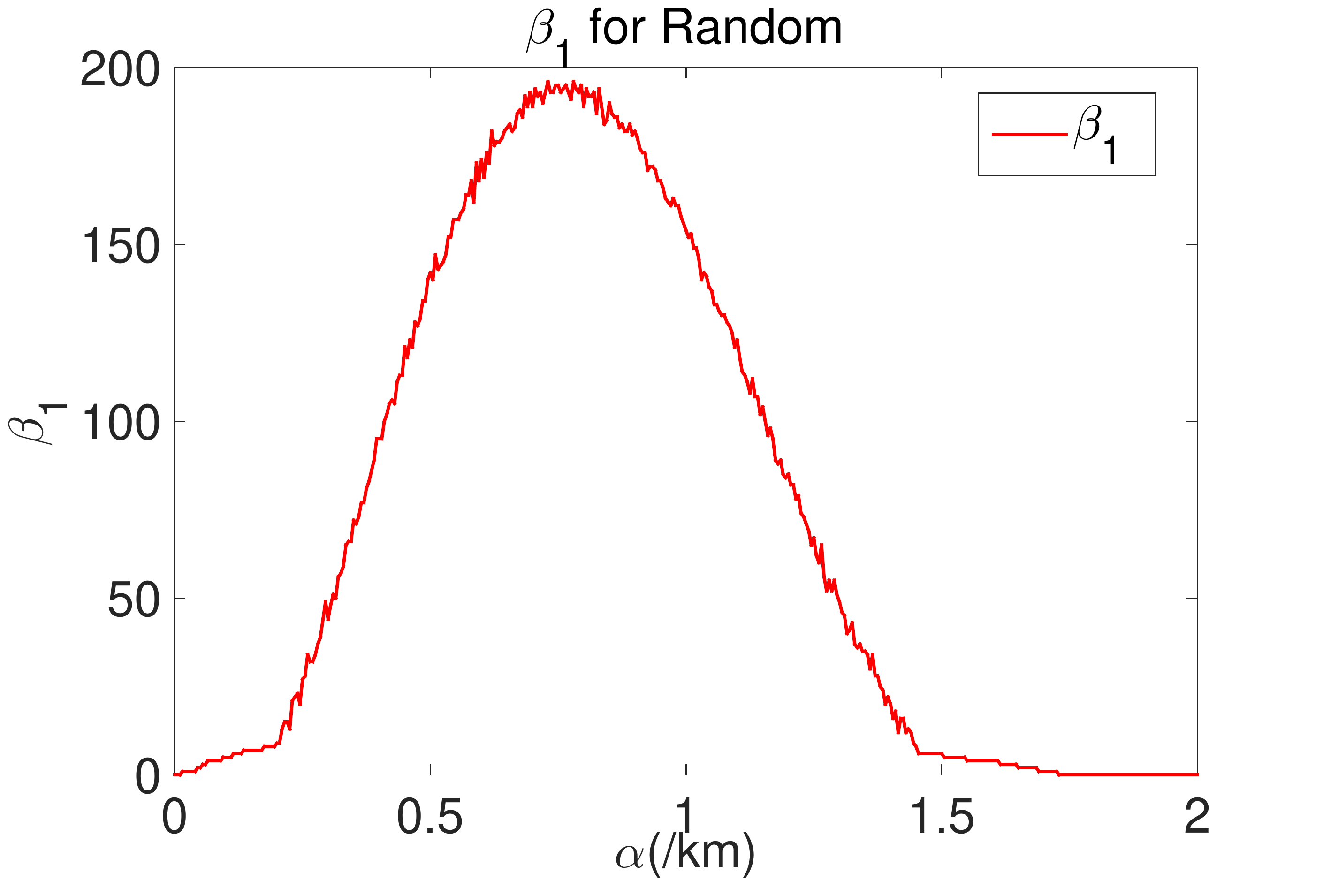}
		}
	}

    \subfigure[Betti curves for fractal: (left) $ \beta_{0} $; (right) $ \beta_{1} $.]{
	\makebox[4cm][c]{
		\includegraphics[scale=0.13]{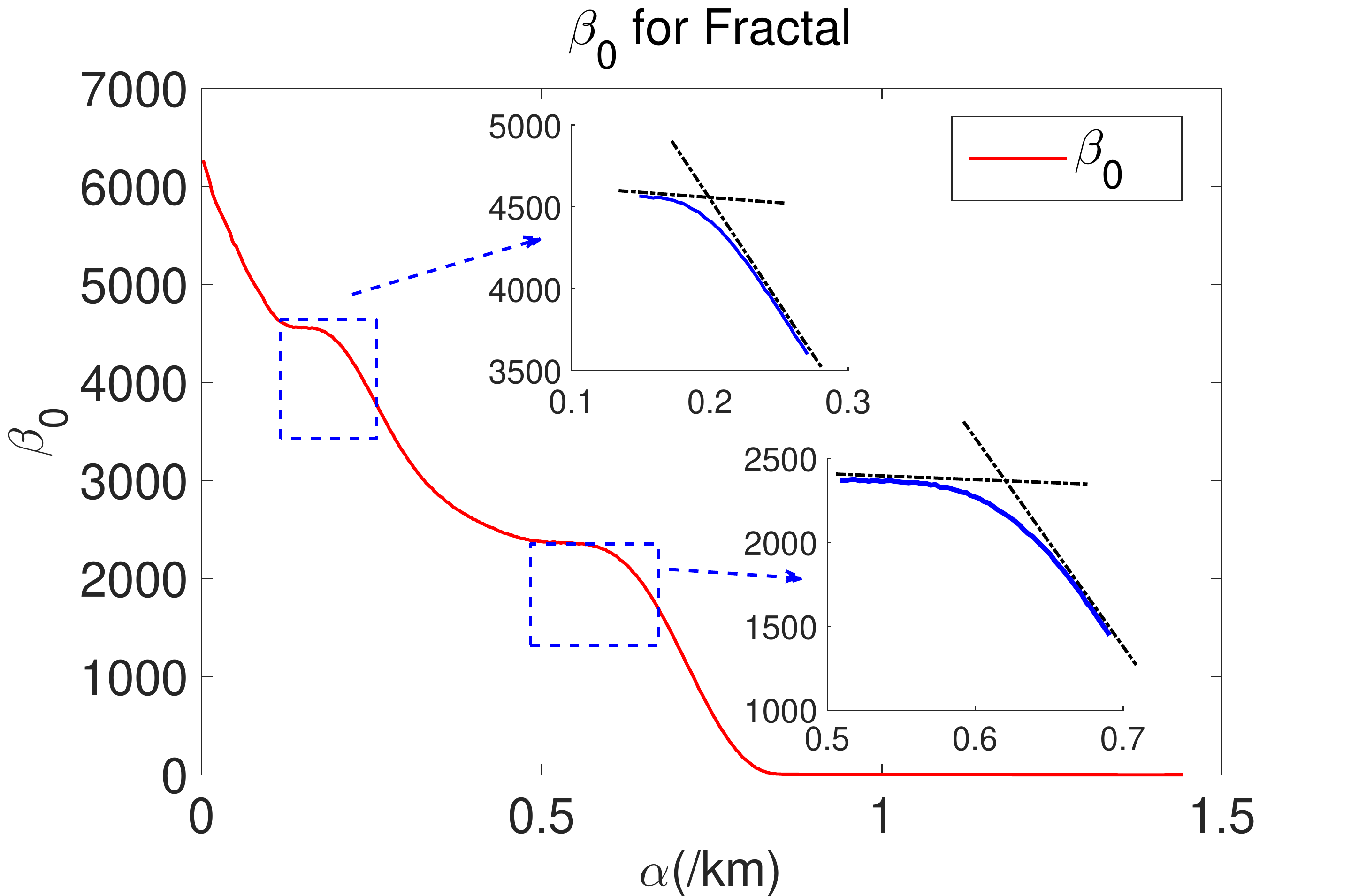}
	}
	\makebox[4cm][c]{
		\includegraphics[scale=0.13]{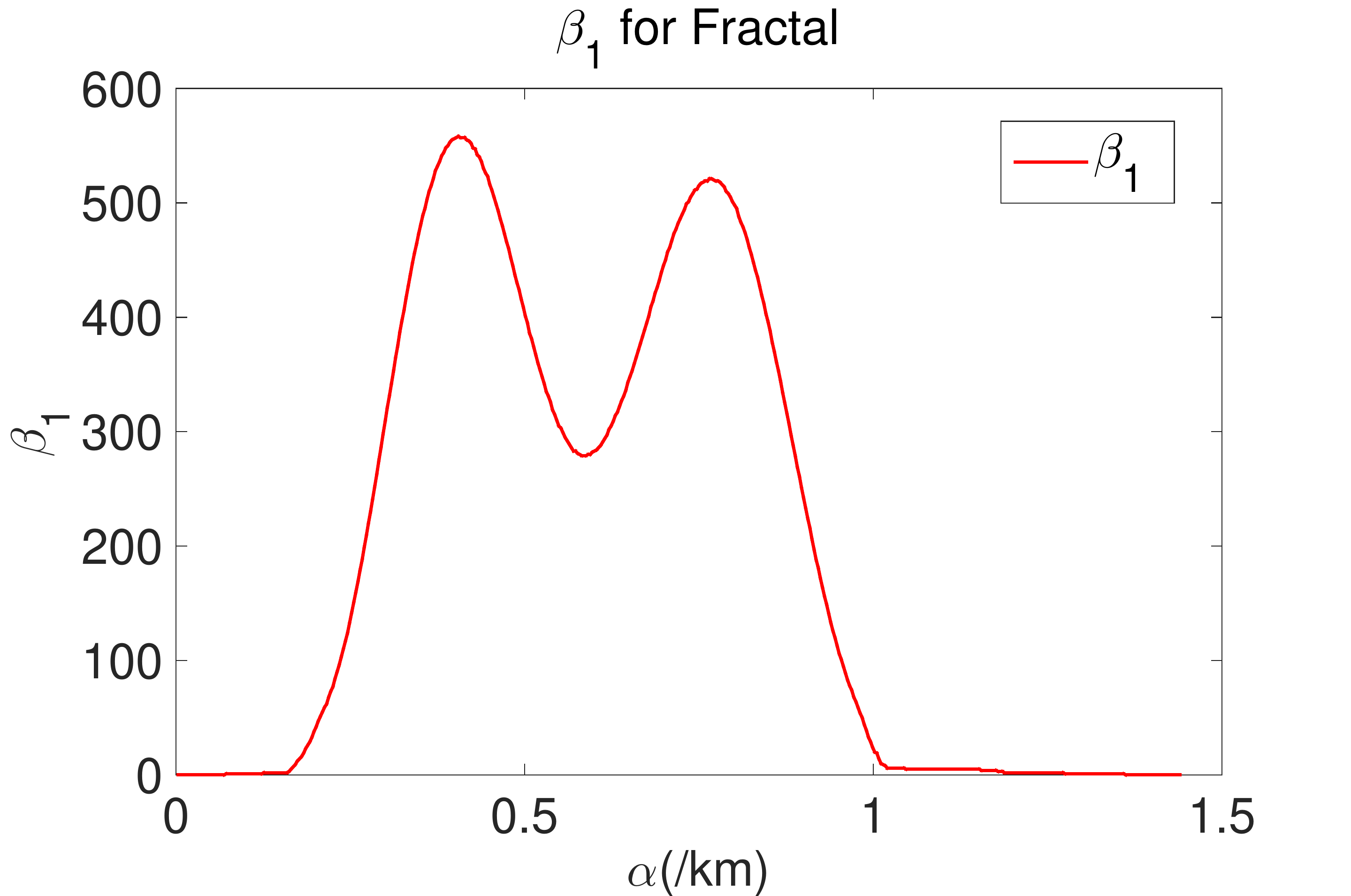}
	}
}
	
	\caption{Comparison between the Betti curves of random and fractal point distributions. }
\end{figure}

Fig. 1(a) displays the practical point diagrams for both the random and fractal point deployments. For the left part of Fig. 1(a), the horizontal and vertical coordinates of each point are randomly designated according to Poisson point distribution, whereas the fractal point deployment in the right part of Fig. 1(a) is realized by the hierarchical partitions of the area, which brings the most principal aspect of fractal features, i.e., self-similarity \cite{Benhaiem2018Self}, into the point distribution. 

Fig. 1(b) illustrates the Betti curves for the random pattern, while Fig. 1(c) for the fractal pattern. Instead of the clearly monotonous decease in the $ \beta_{0} $ curve and the single peak in the $ \beta_{1} $ curve in the random situation, the fractal nature is manifested by the distinctive features of multiple ripples and peaks in the $ \beta_{0} $ and $ \beta_{1} $ curves, respectively \cite{Pranav2017The}, where a ripple is formed due to the distinct slope change as highlighted by the amplified blue subgraph in Fig. 1(c).  

In summary, the fractal property can be characterized by the hierarchical patterns, i.e., the multiple ripples and peaks of the Betti curves \cite{Pranav2017The}. In our works, the fractal property in the BS topology is verified in Fig. 2 for European cities and Fig. 3 for Asian ones, respectively. Beyond the geographical constraints, it is extremely astonishing to find out the consistent fractal nature for all the aforementioned eight cities.

\begin{figure*}[htbp]
	\centering
	\subfigure[London$ \qquad\qquad\qquad\qquad\qquad\qquad\qquad\qquad\qquad\qquad\qquad\qquad $(c) Paris]{
		\makebox[4cm][c]{
			\includegraphics[scale=0.13]{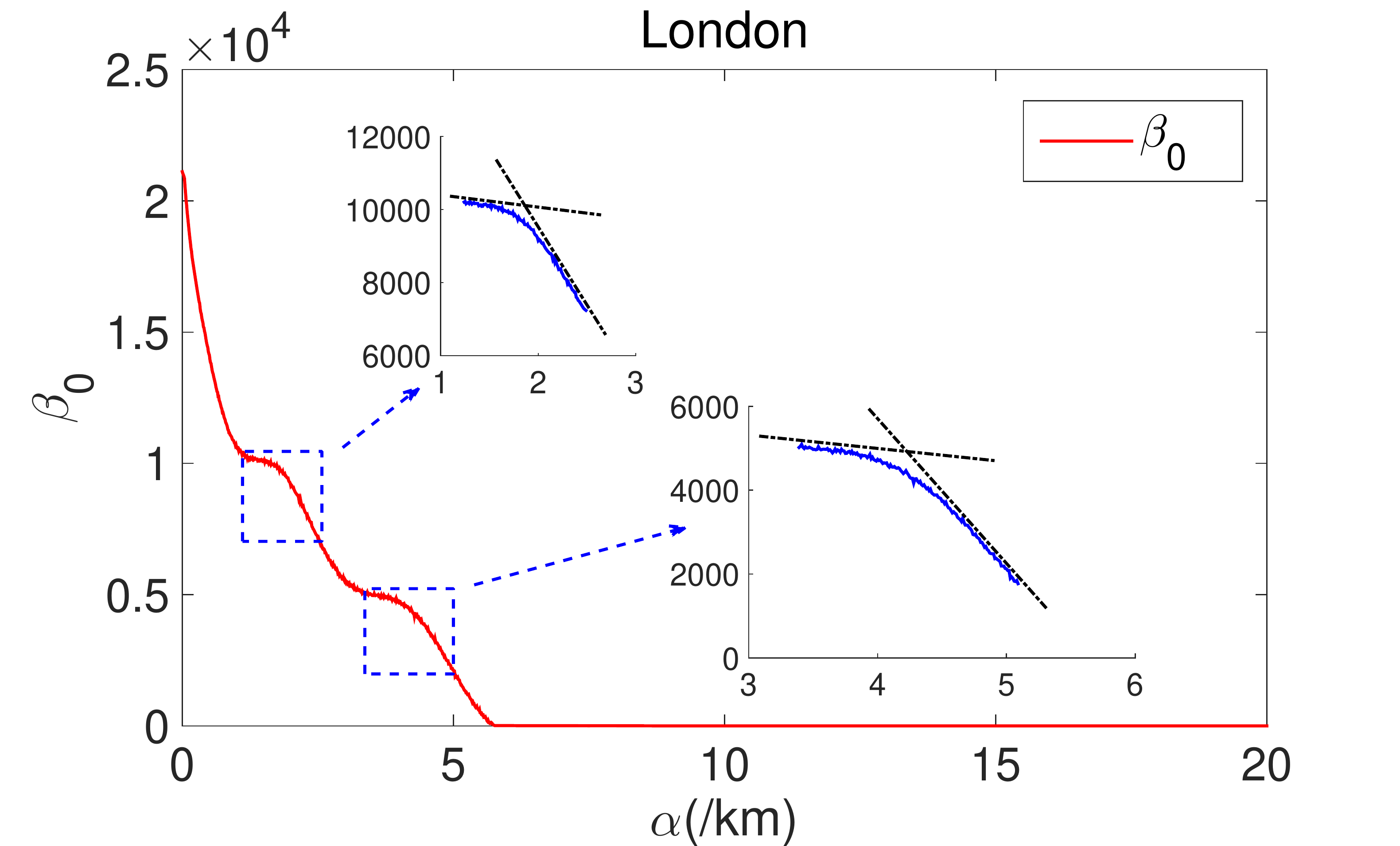}
		}
		\makebox[4cm][c]{
			\includegraphics[scale=0.13]{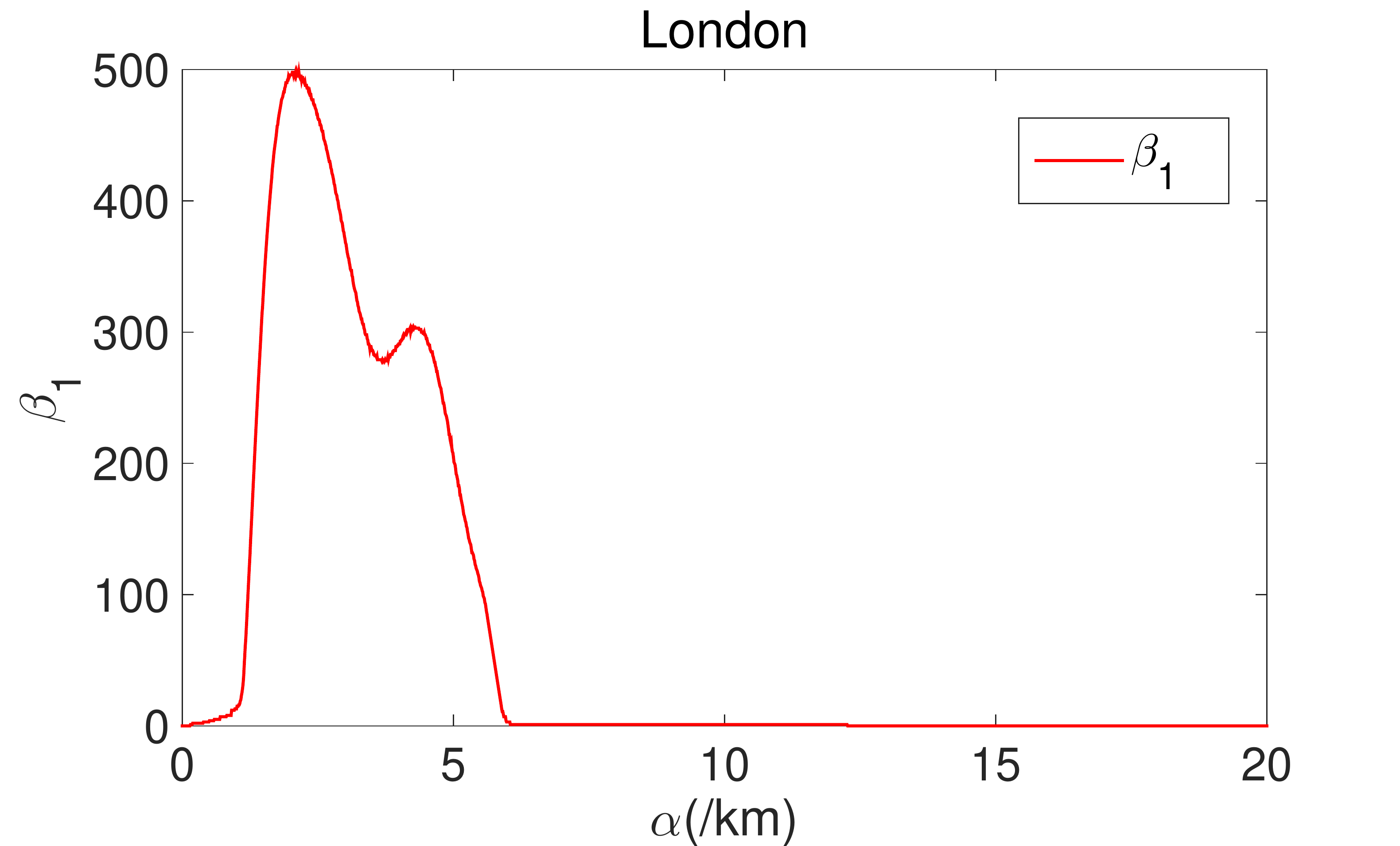}
		}
	   \makebox[4cm][c]{
		\includegraphics[scale=0.13]{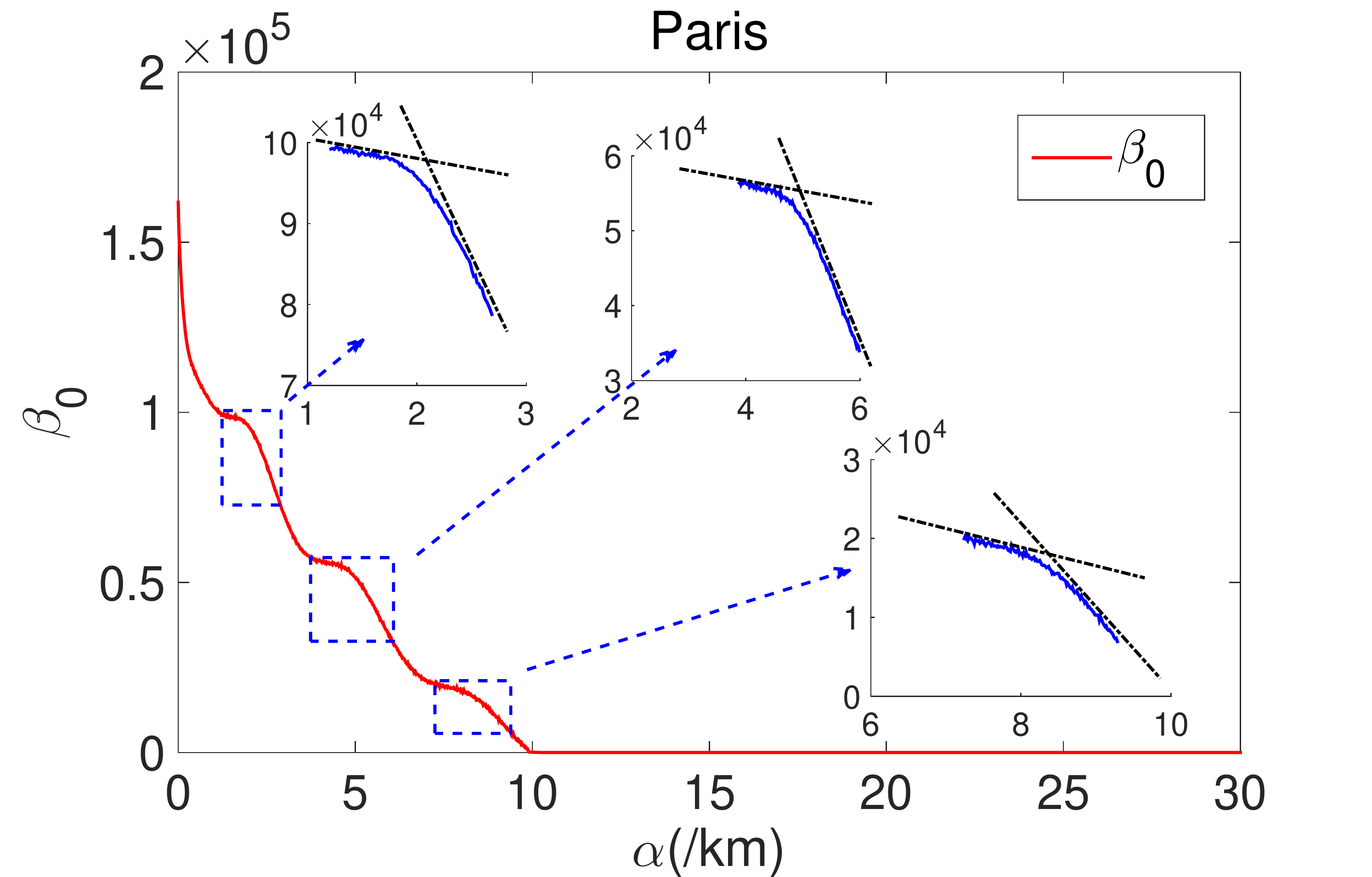}
	}
	   \makebox[4cm][c]{
		\includegraphics[scale=0.13]{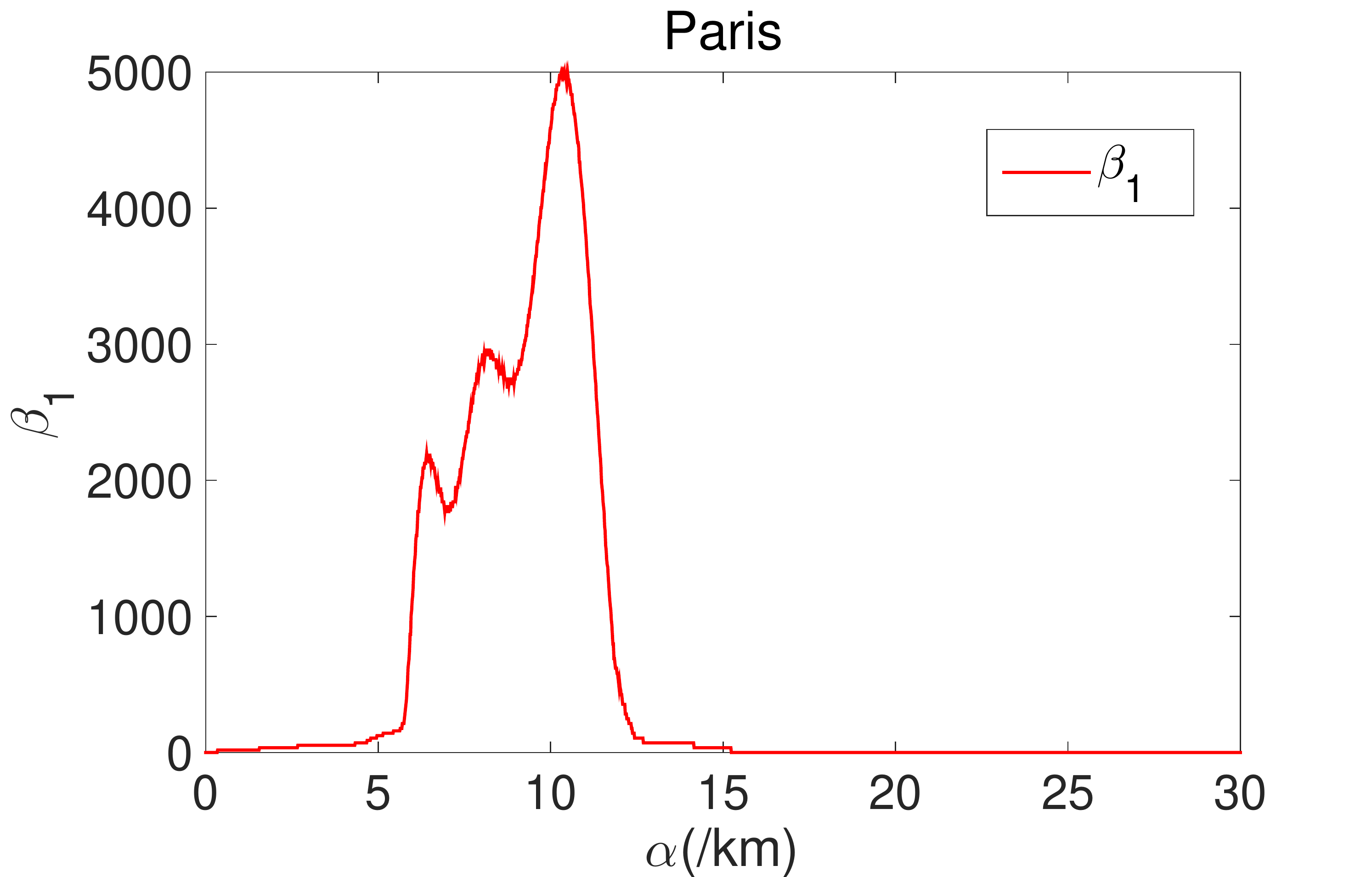}
	}
}
	
	\subfigure[Munich$ \qquad\qquad\qquad\qquad\qquad\qquad\qquad\qquad\qquad\qquad\qquad\qquad $(d) Warsaw]{
		\makebox[4cm][c]{
			\includegraphics[scale=0.13]{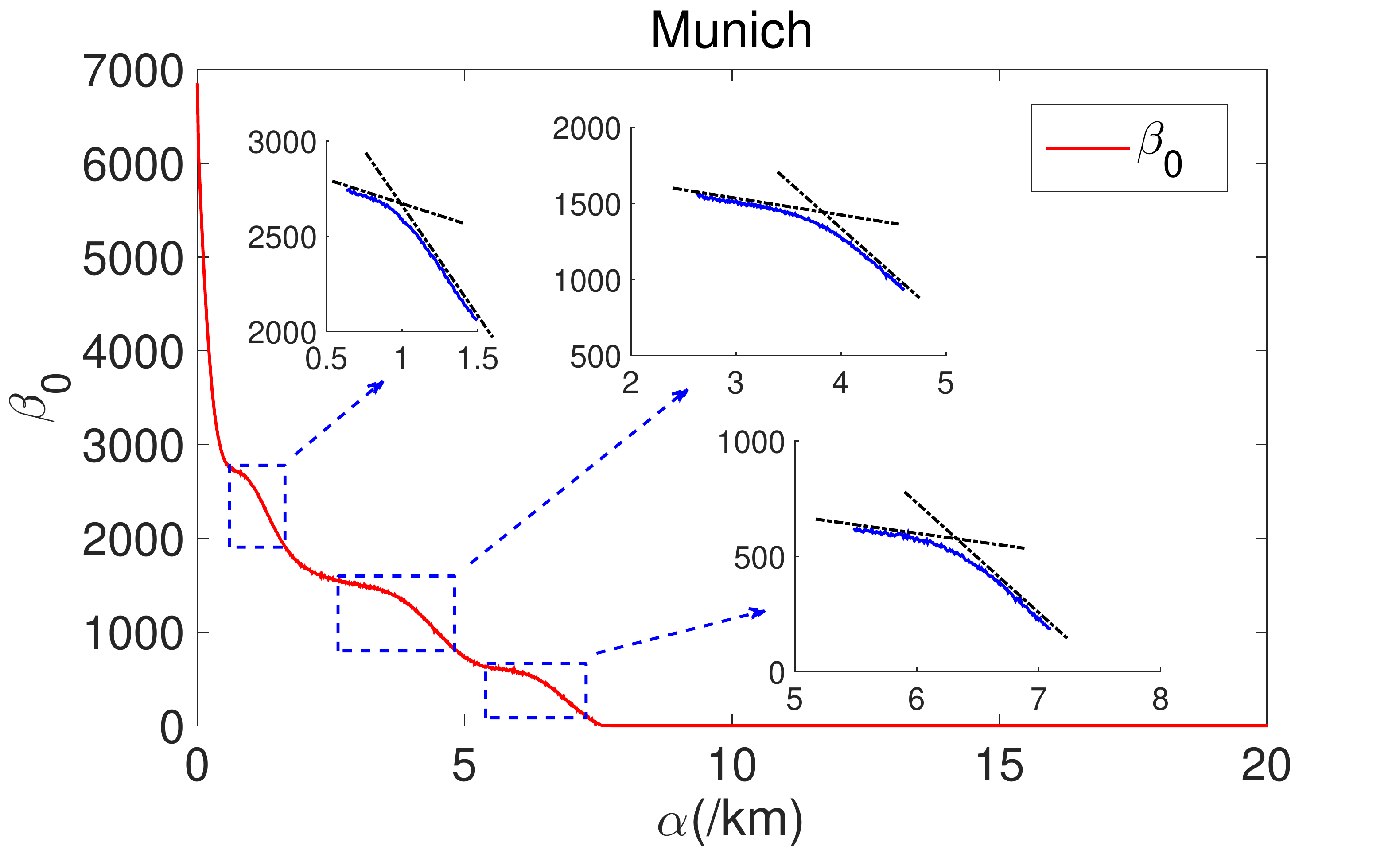}
		}
		\makebox[4cm][c]{
			\includegraphics[scale=0.13]{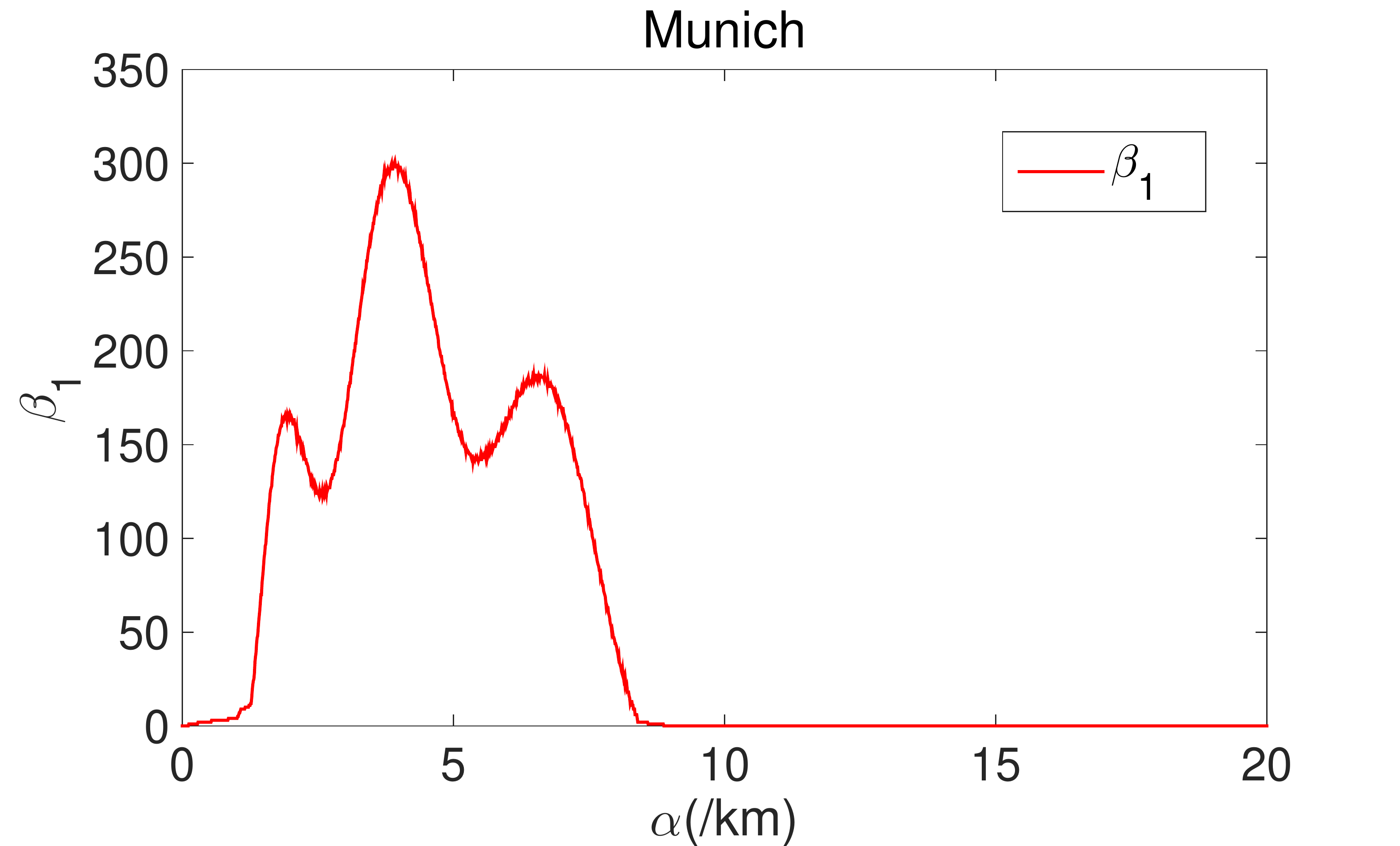}
		}
		\makebox[4cm][c]{
		\includegraphics[scale=0.13]{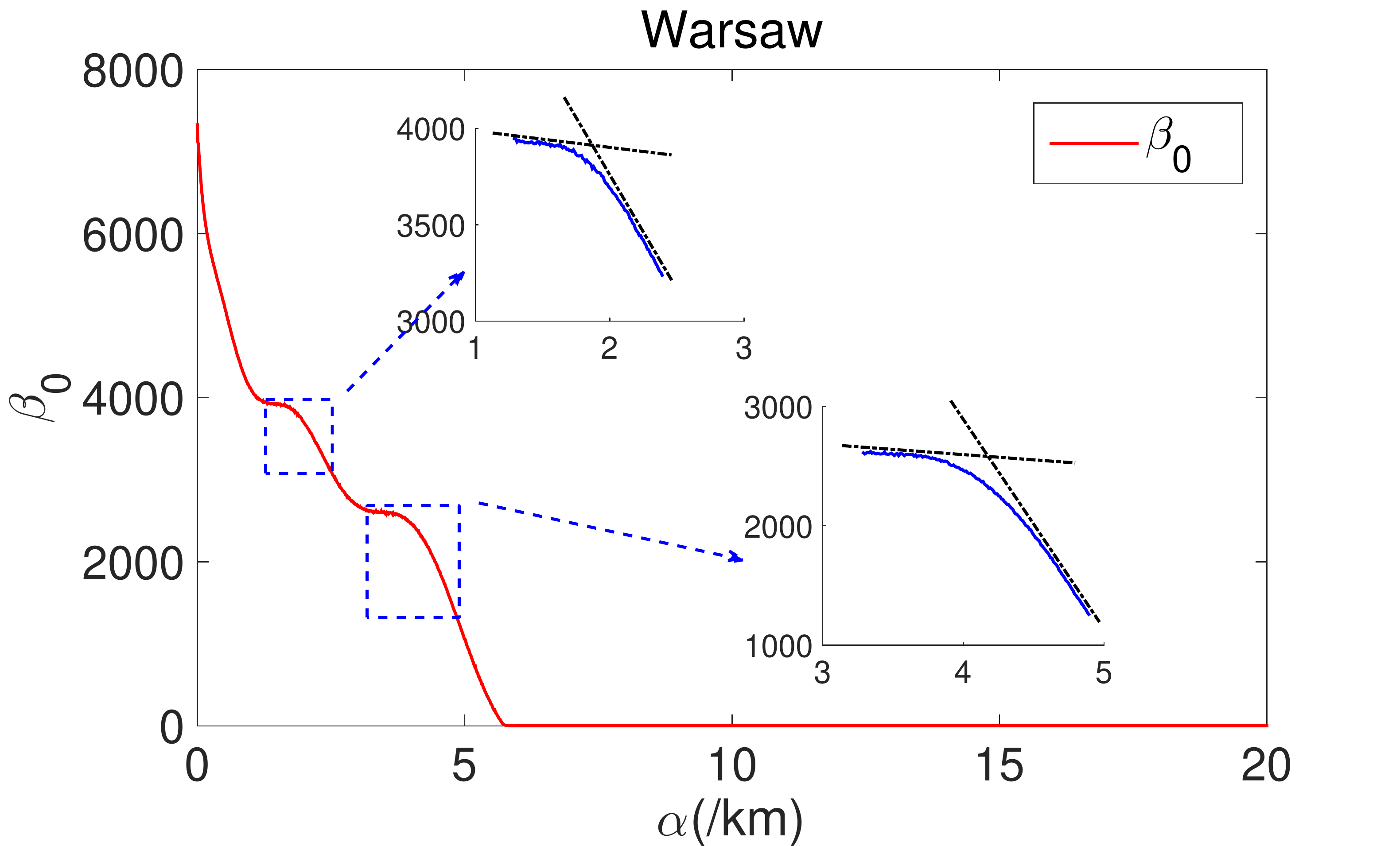}
	}
	    \makebox[4cm][c]{
		\includegraphics[scale=0.13]{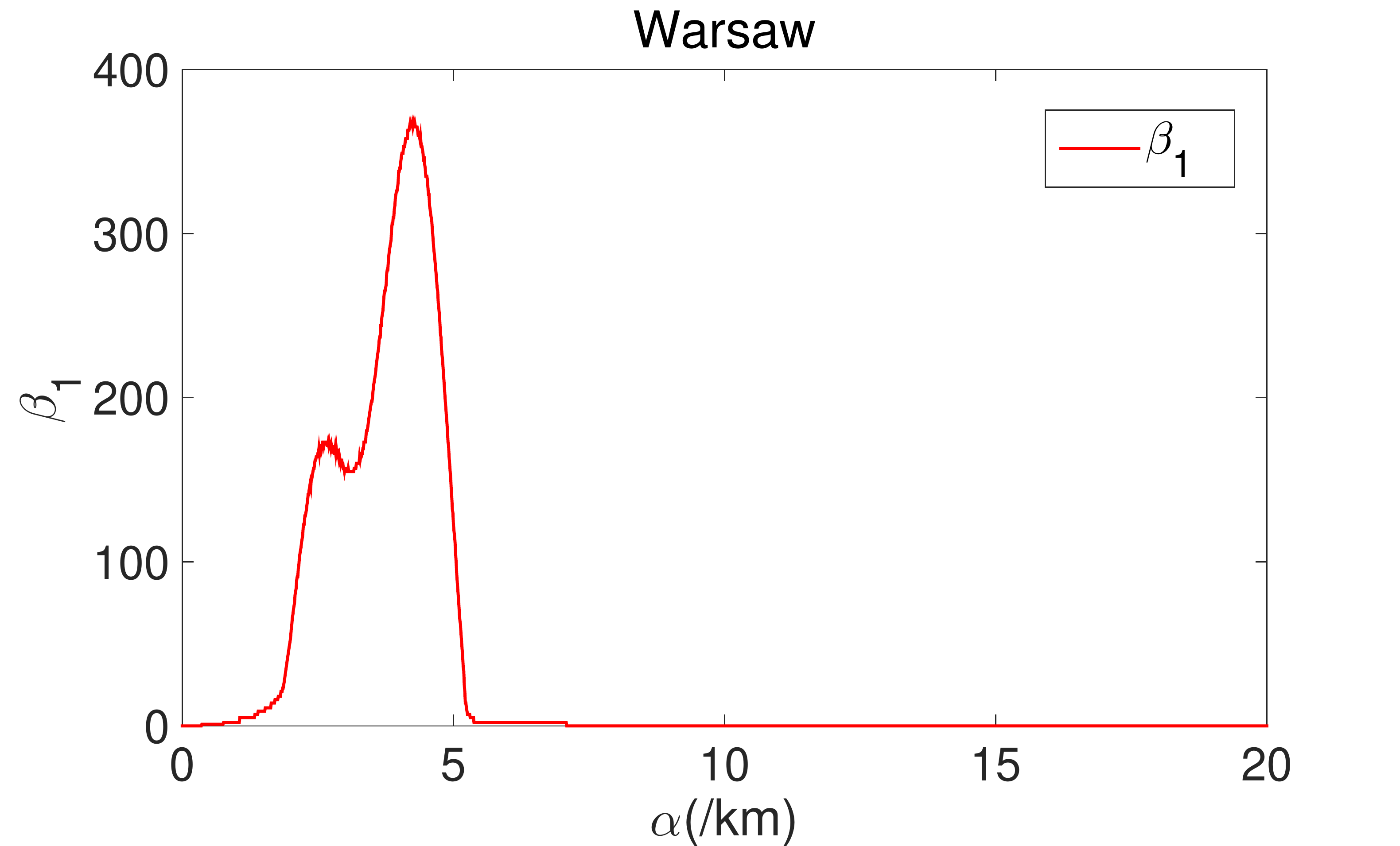}
	}
}

	\caption{The Betti curves of the practical BS deployments in European cities.}
\end{figure*}

\begin{figure*}[htbp]
	\centering
	\subfigure[Beijing$ \qquad\qquad\qquad\qquad\qquad\qquad\qquad\qquad\qquad\qquad\qquad\qquad $(c) Seoul]{
		\makebox[4cm][c]{
			\includegraphics[scale=0.13]{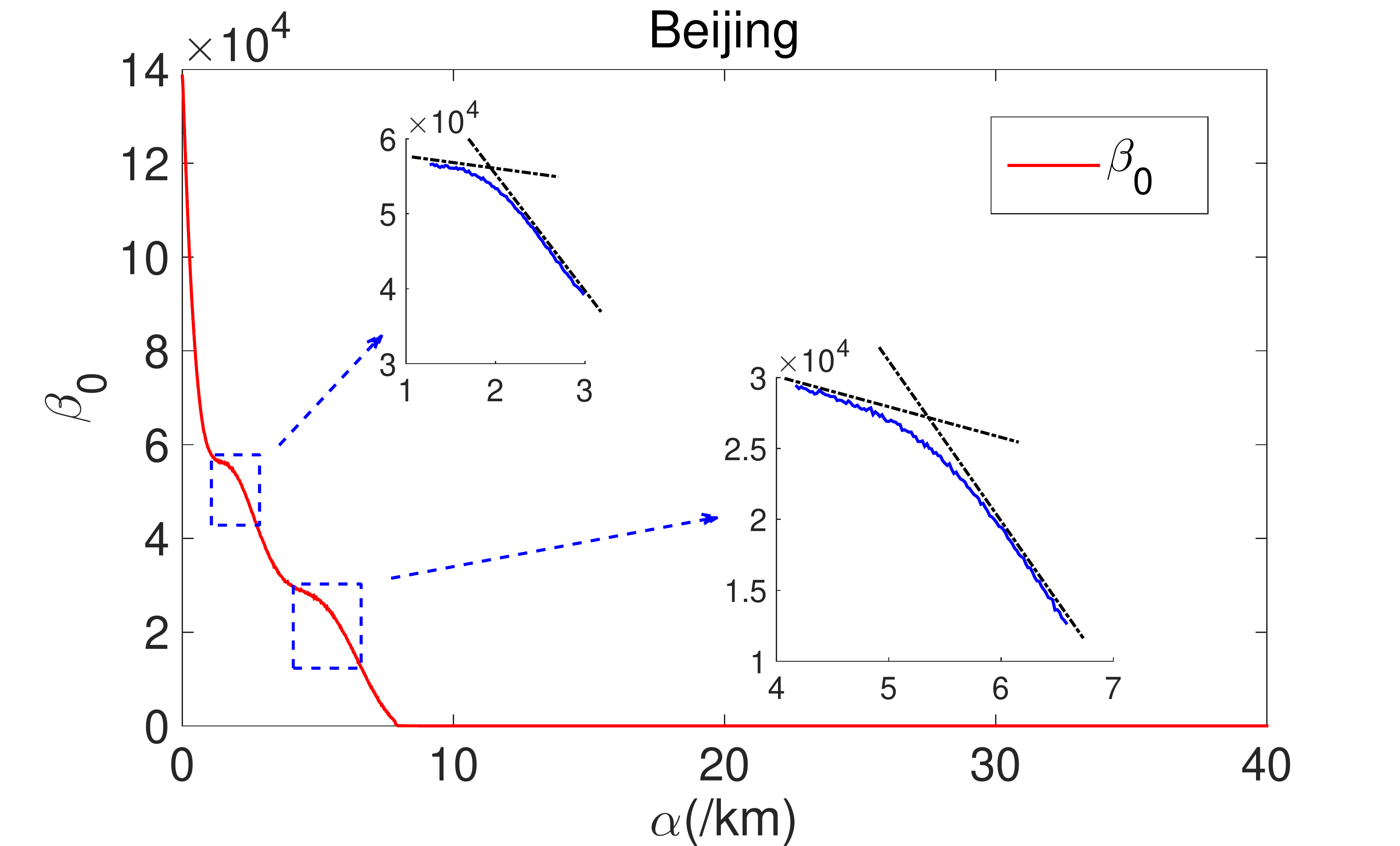}
		}
		\makebox[4cm][c]{
			\includegraphics[scale=0.13]{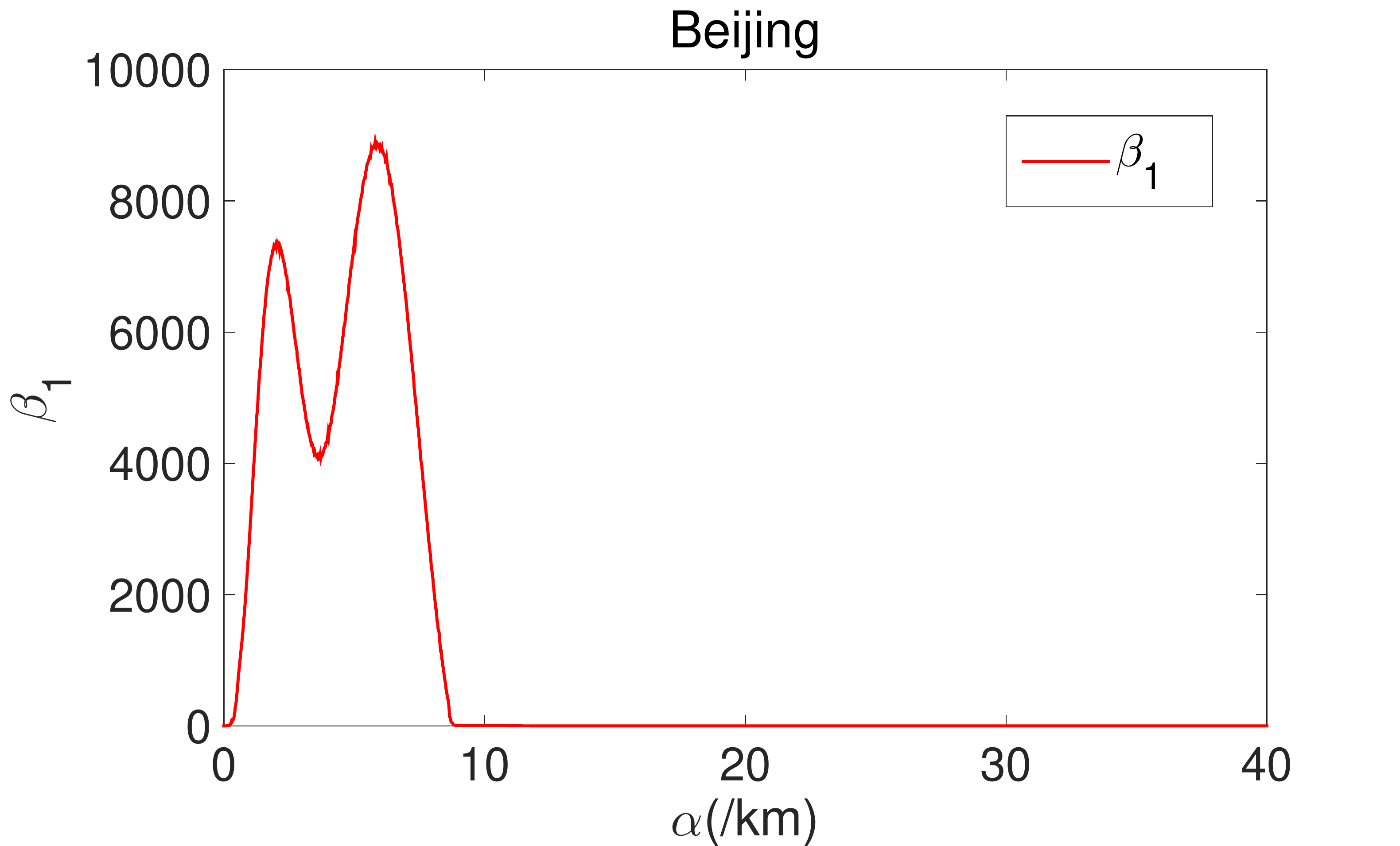}
		}
		\makebox[4cm][c]{
			\includegraphics[scale=0.13]{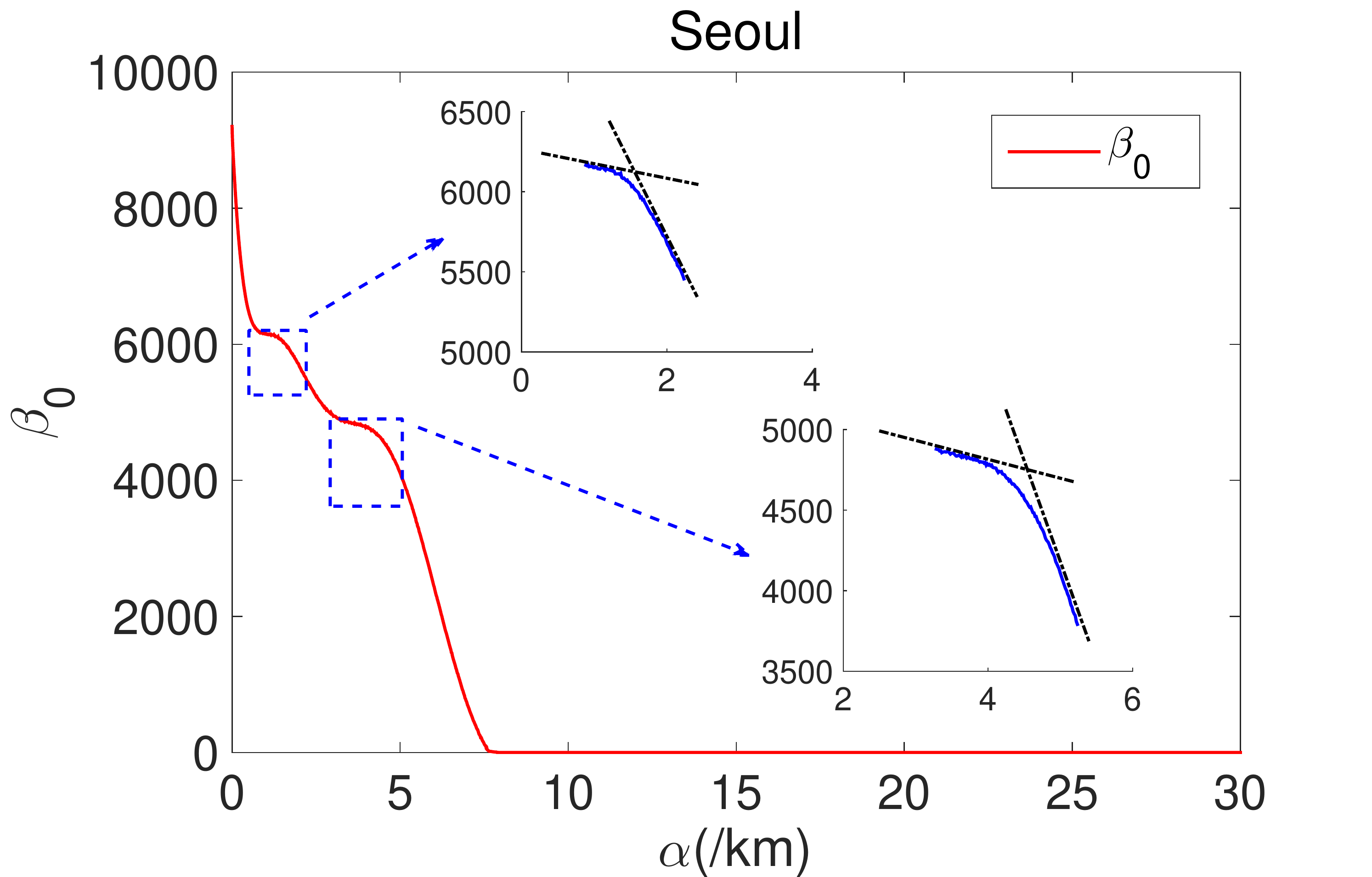}
		}
		\makebox[4cm][c]{
			\includegraphics[scale=0.13]{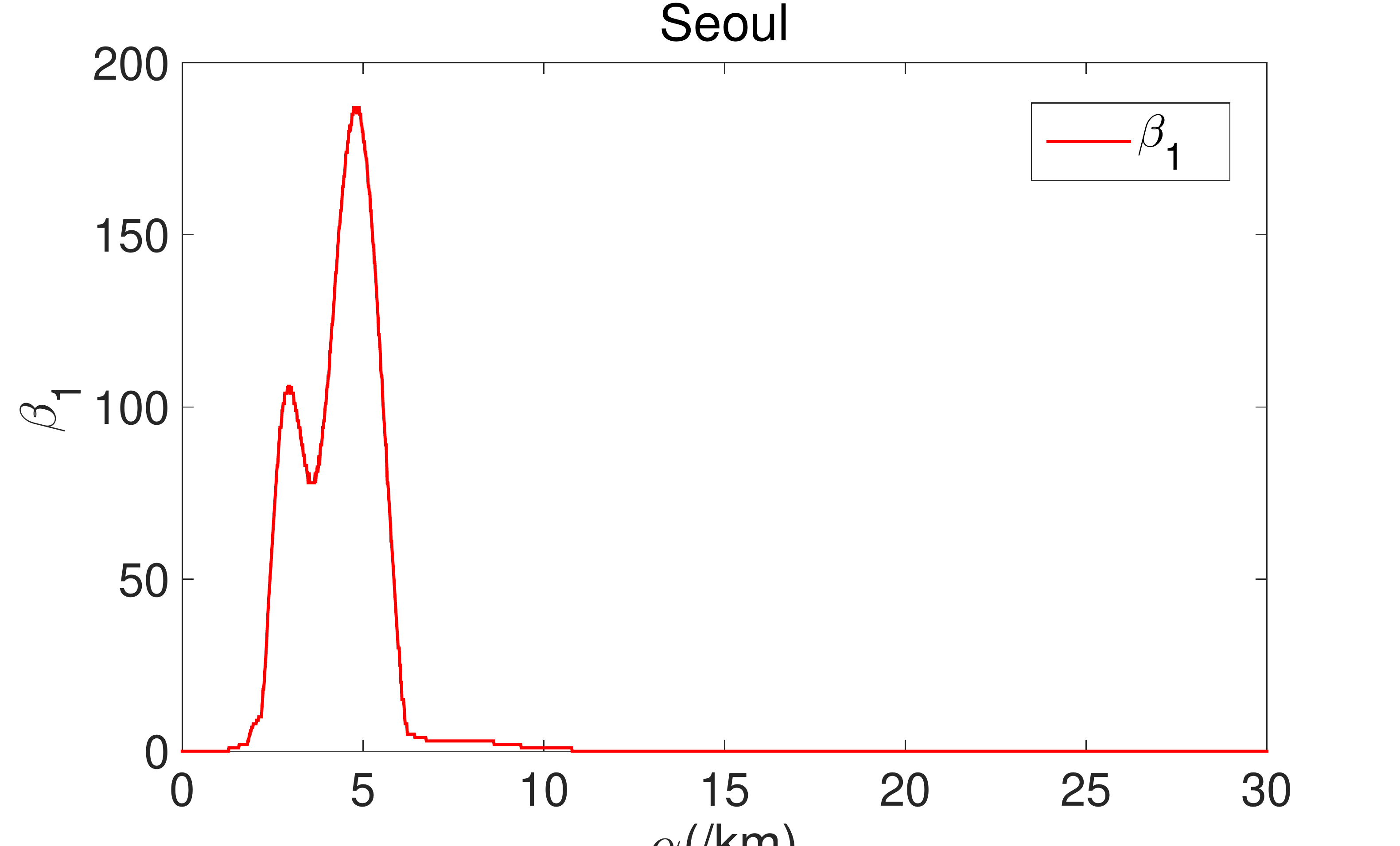}
		}
	}
	
	\subfigure[Mumbai$ \qquad\qquad\qquad\qquad\qquad\qquad\qquad\qquad\qquad\qquad\qquad\qquad $(d) Tokyo]{
		\makebox[4cm][c]{
			\includegraphics[scale=0.13]{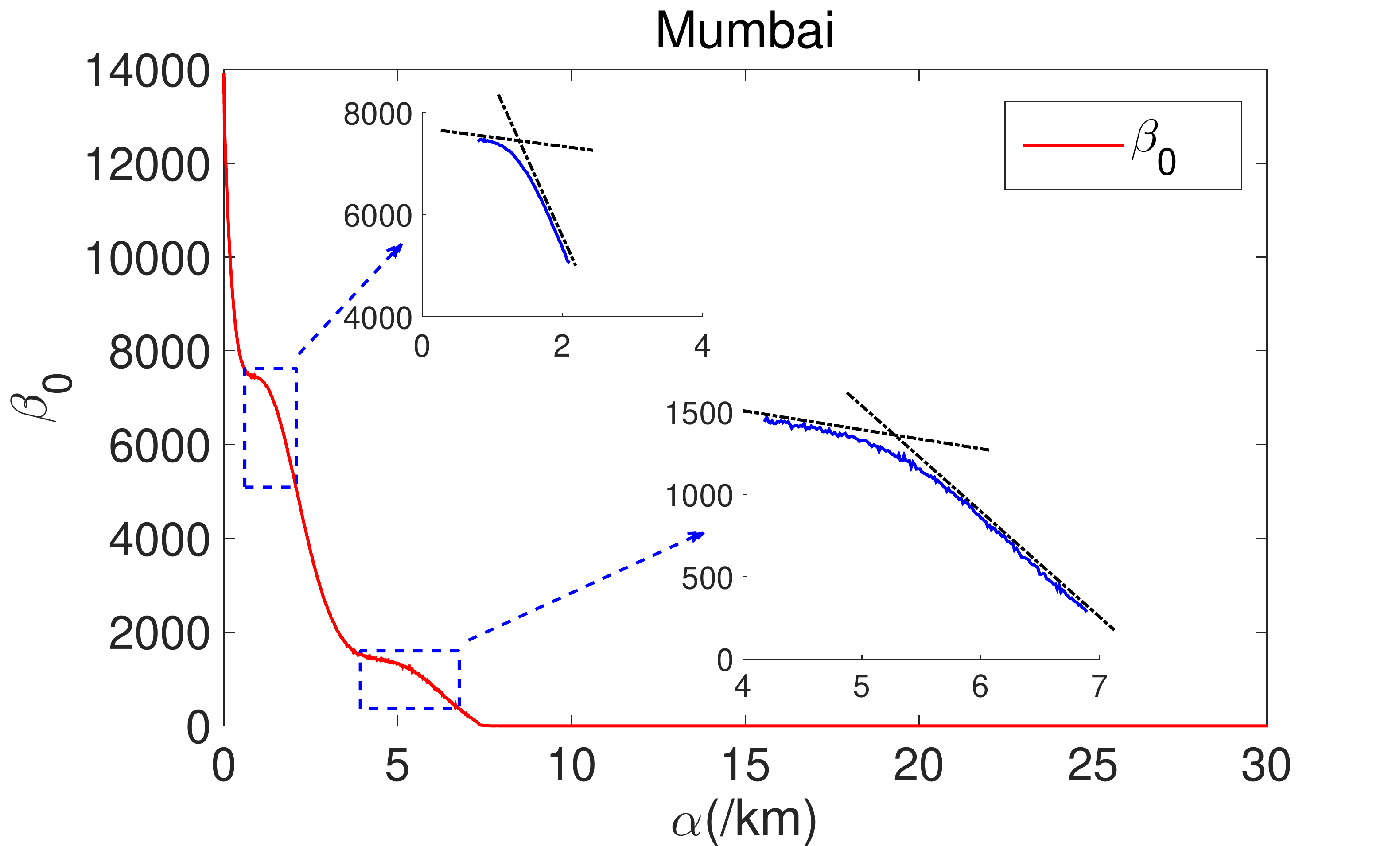}
		}
		\makebox[4cm][c]{
			\includegraphics[scale=0.13]{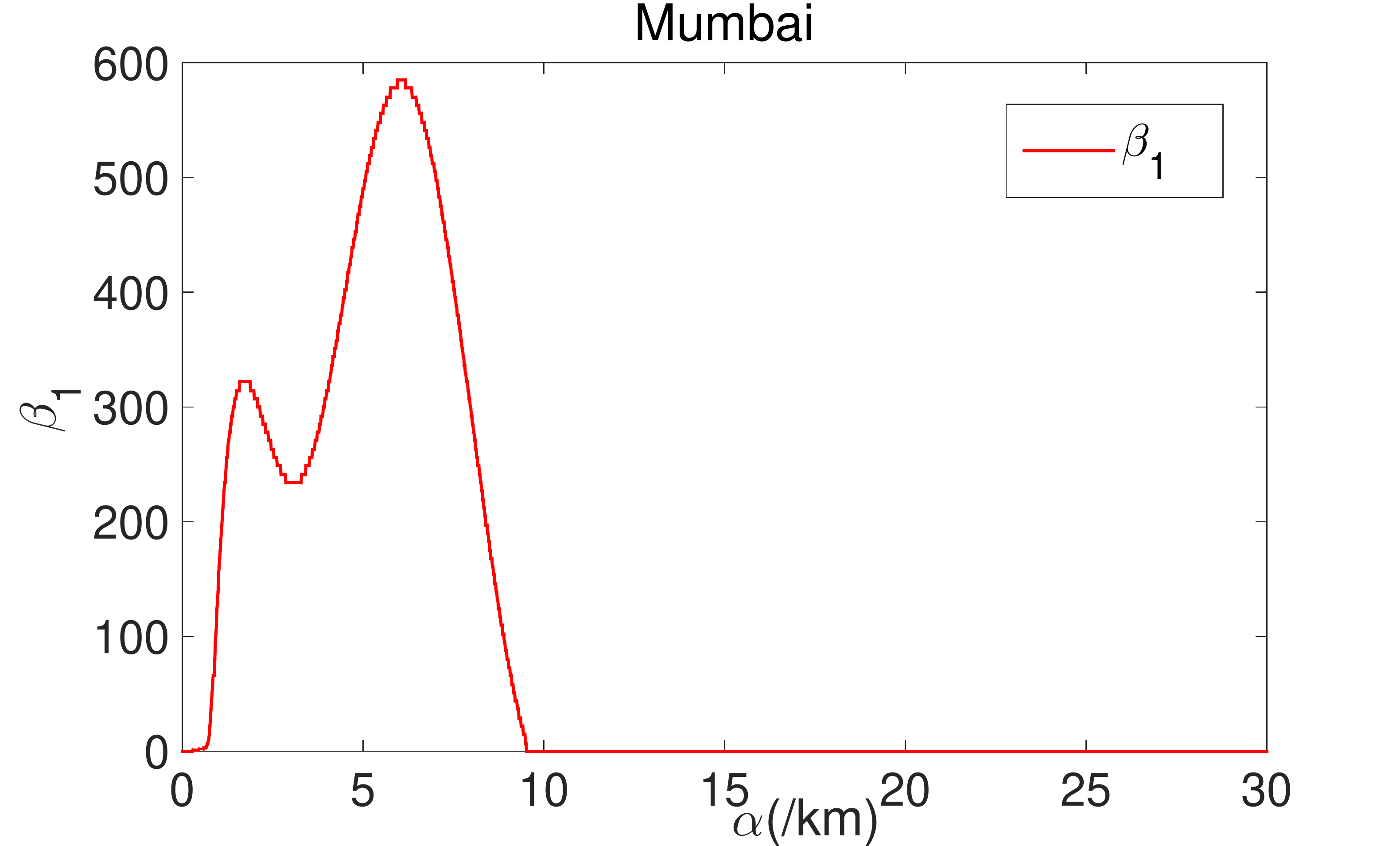}
		}
		\makebox[4cm][c]{
			\includegraphics[scale=0.13]{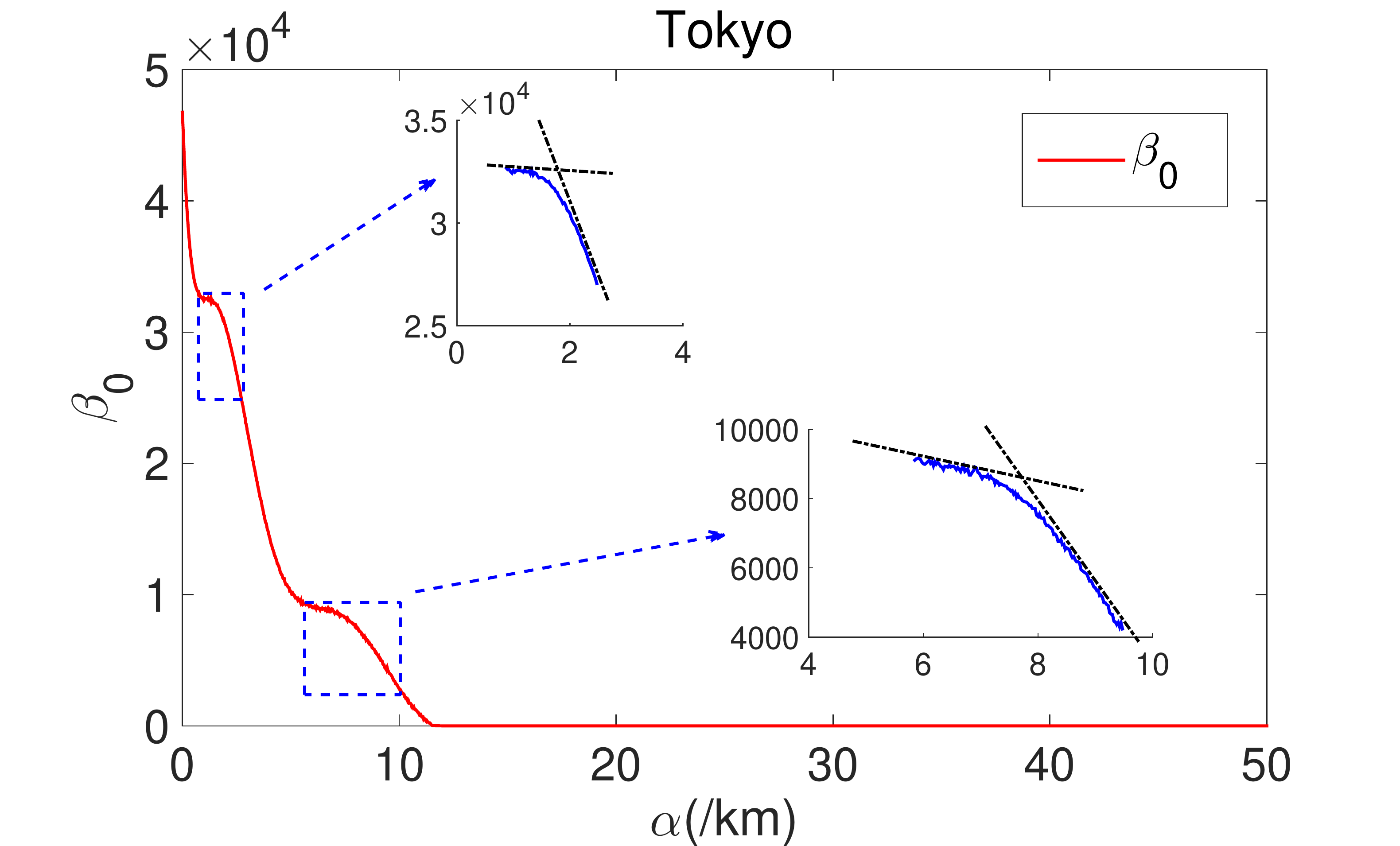}
		}
		\makebox[4cm][c]{
			\includegraphics[scale=0.13]{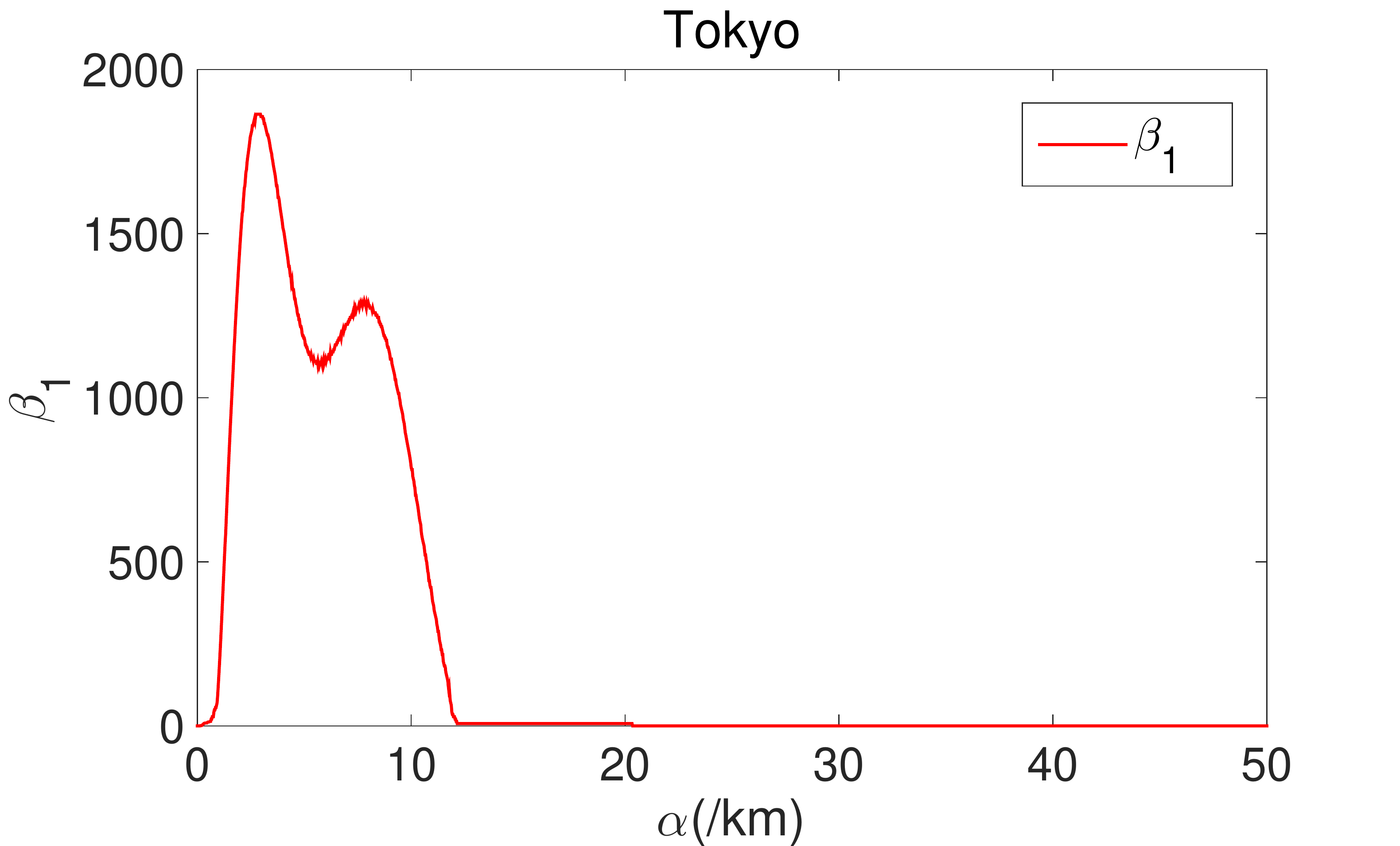}
		}
	}

	\caption{The Betti curves of the practical BS deployments in Asian cities.}
\end{figure*}

Moreover, the entirely different positions of the multiple ripples and peaks can be evaluated on the basis of the crossover point of two straight lines with quite different gradients, as shown by the $ \beta_{0} $ curves in Fig. 2 and Fig. 3. As listed in Table II, it is clear that the peaks always come after the corresponding ripples, which makes sense because of the larger size of loops than that of components in the relevant topologies.
\begin{table}[htb]
	\centering
	\includegraphics[scale=0.25]{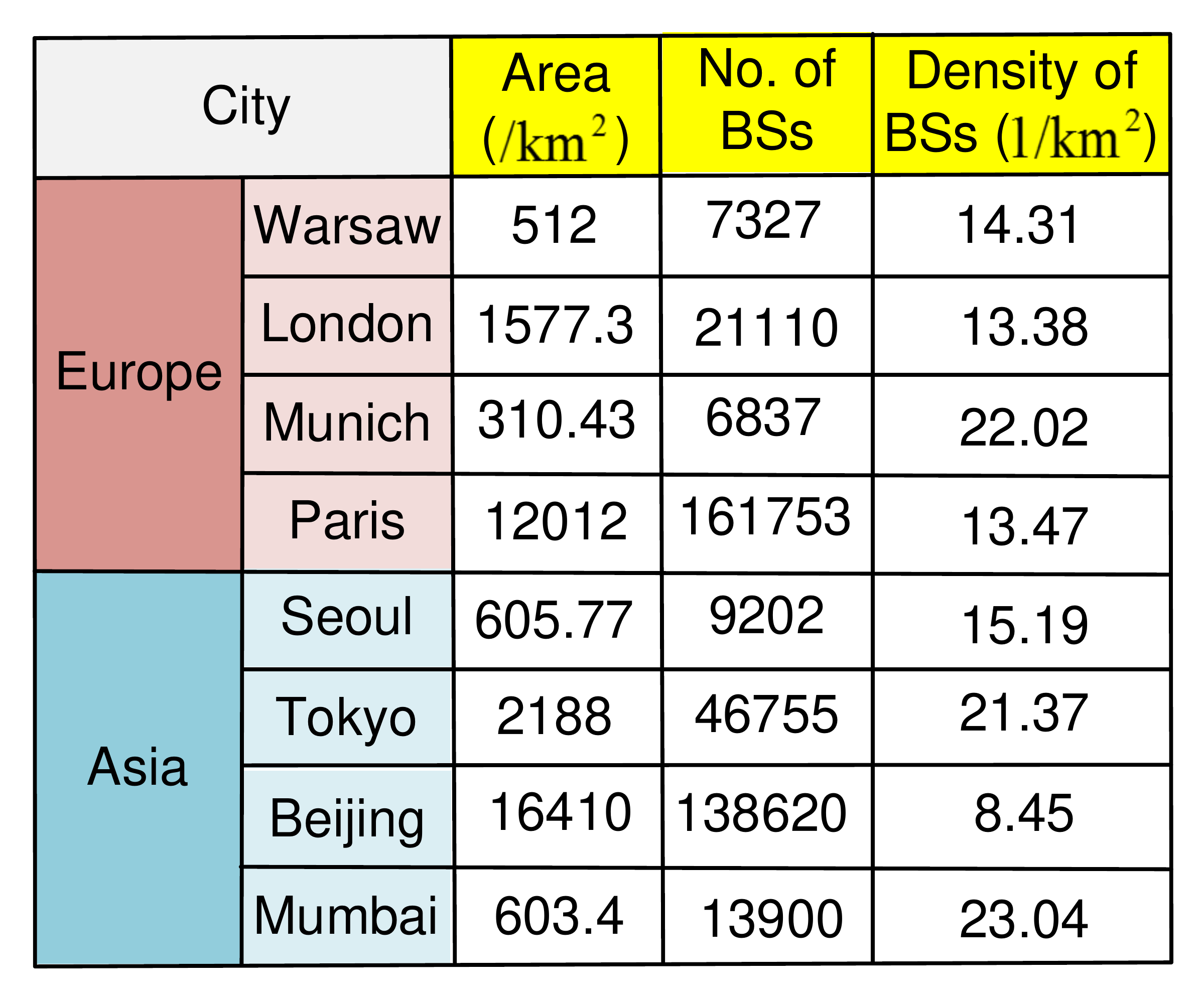}
	\caption{The basic information of 8 selected cities.}
\end{table}

\begin{table}[htb]
	\centering
	\includegraphics[scale=0.4]{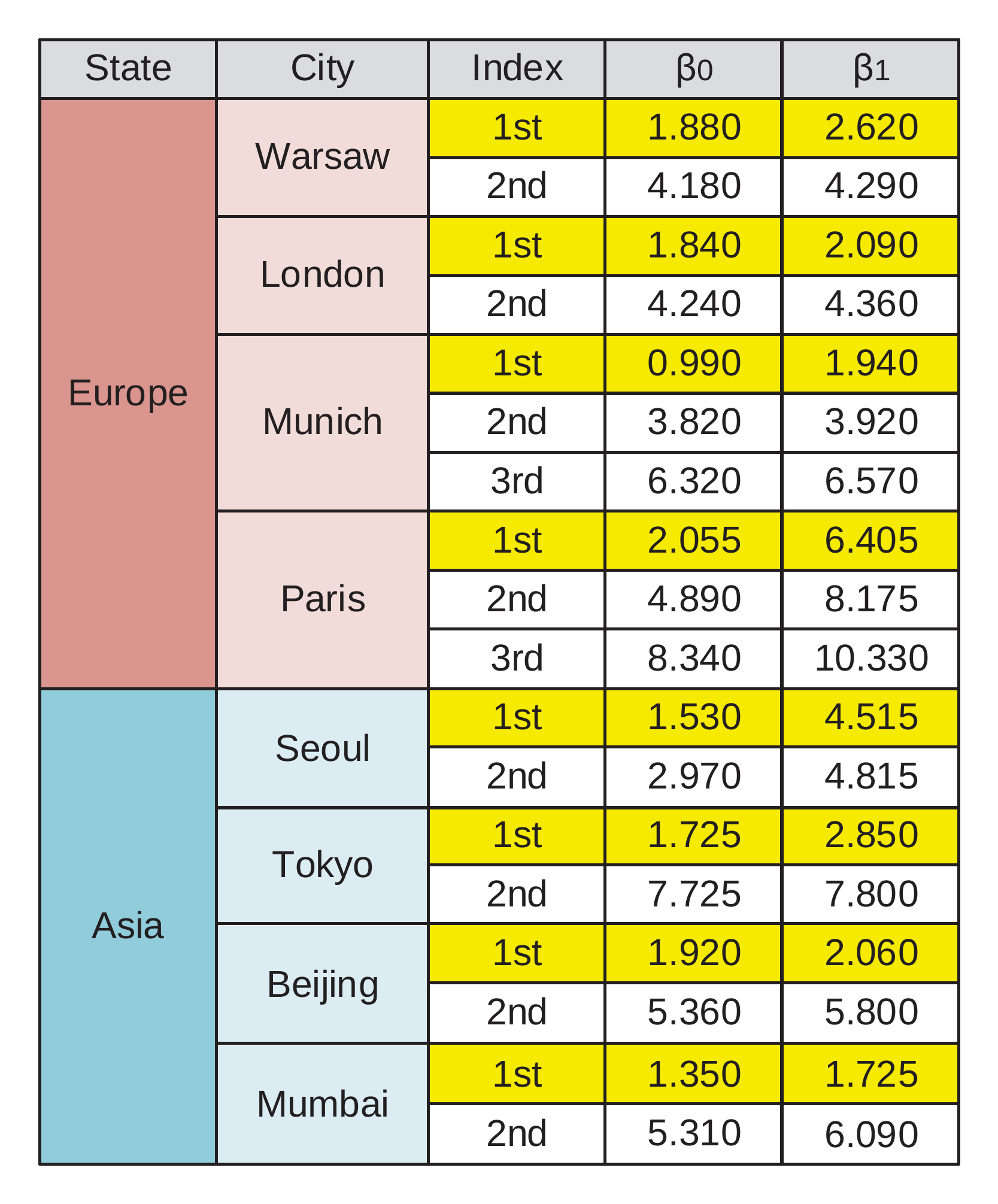}
	\caption{The positions of the ripples and peaks in the Betti curves.}
\end{table}

\subsection{Fractal Features Based on the Hurst Coefficients}
The Hurst coefficient between [0,1] is widely used as an evaluation index of the fractal property of data series, and the indication of fractality increases gradually as the Hurst coefficient approaches to 1 \cite{Gneiting2001Stochastic}. 

For the sake of verifying the fractal nature in the BS topology, the rescaled range analysis (R/S) method \cite{Fern2014An} is adopted to compute the Hurst coefficients of all the selected cities in this letter. Due to the space limitation, the details about the calculation of the Hurst coefficients will not be provided here, and all the relevant information can be found in \cite{Chen2018Topology}. As listed in Table III, the fractal nature is evidently affirmed because all the Hurst coefficients turn out to be very close to 1.
\begin{table}[htb]
	\centering
	\includegraphics[scale=0.25]{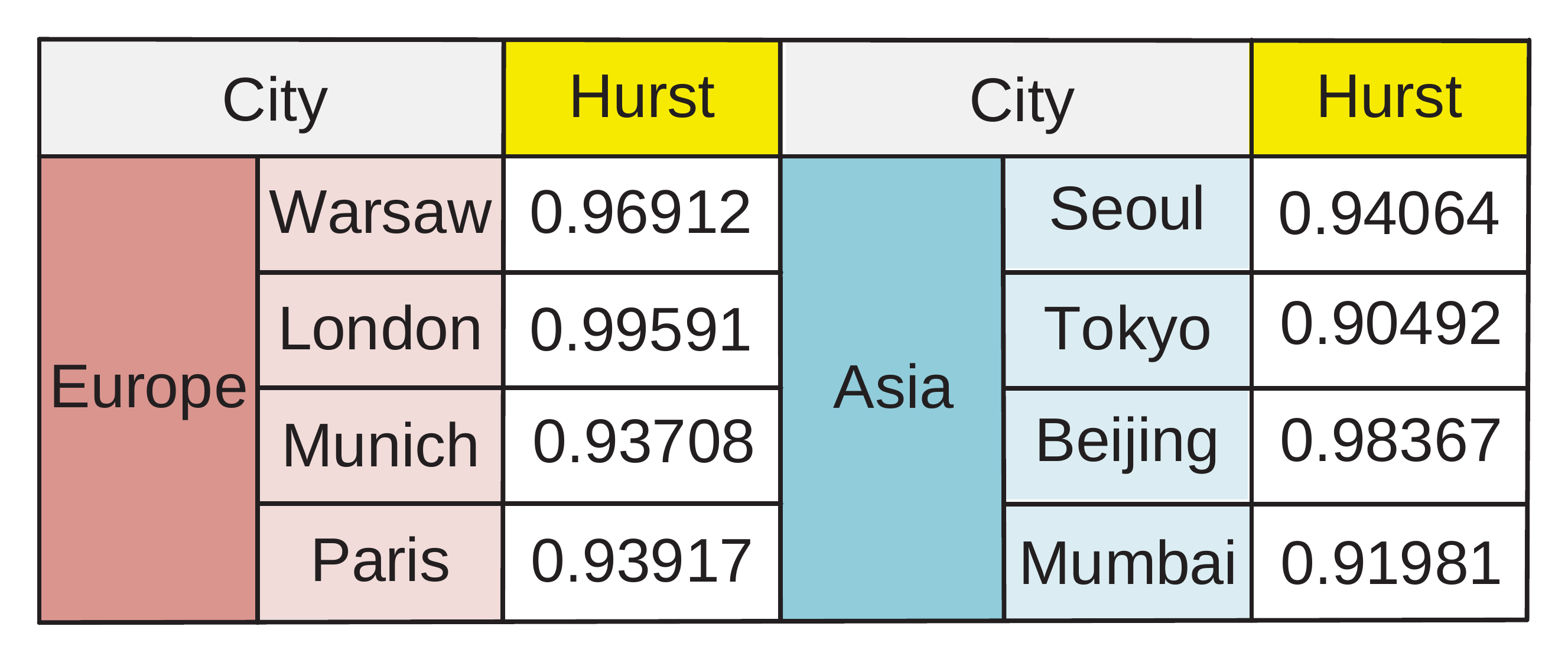}
	\caption{The Hurst coefficients for all the 8 cities.}
\end{table}

\section{Log-normal Distribution of the Euler Characteristics}
The Euler characteristics can be calculated from Betti numbers according to Euler-Poincare Formula  \cite{Weygaert2011Alpha}. Since a clear heavy-tailed property is demonstrated in the probability density functions (PDFs) of the Euler characteristics of the real BS location data, three classical heavy-tailed statistical distributions and widely-used Poisson distribution are selected as the candidates to match the PDFs. The candidates and their PDF formulas are given in Table IV.
\begin{table}[htb]
	\centering
	\includegraphics[scale=0.12]{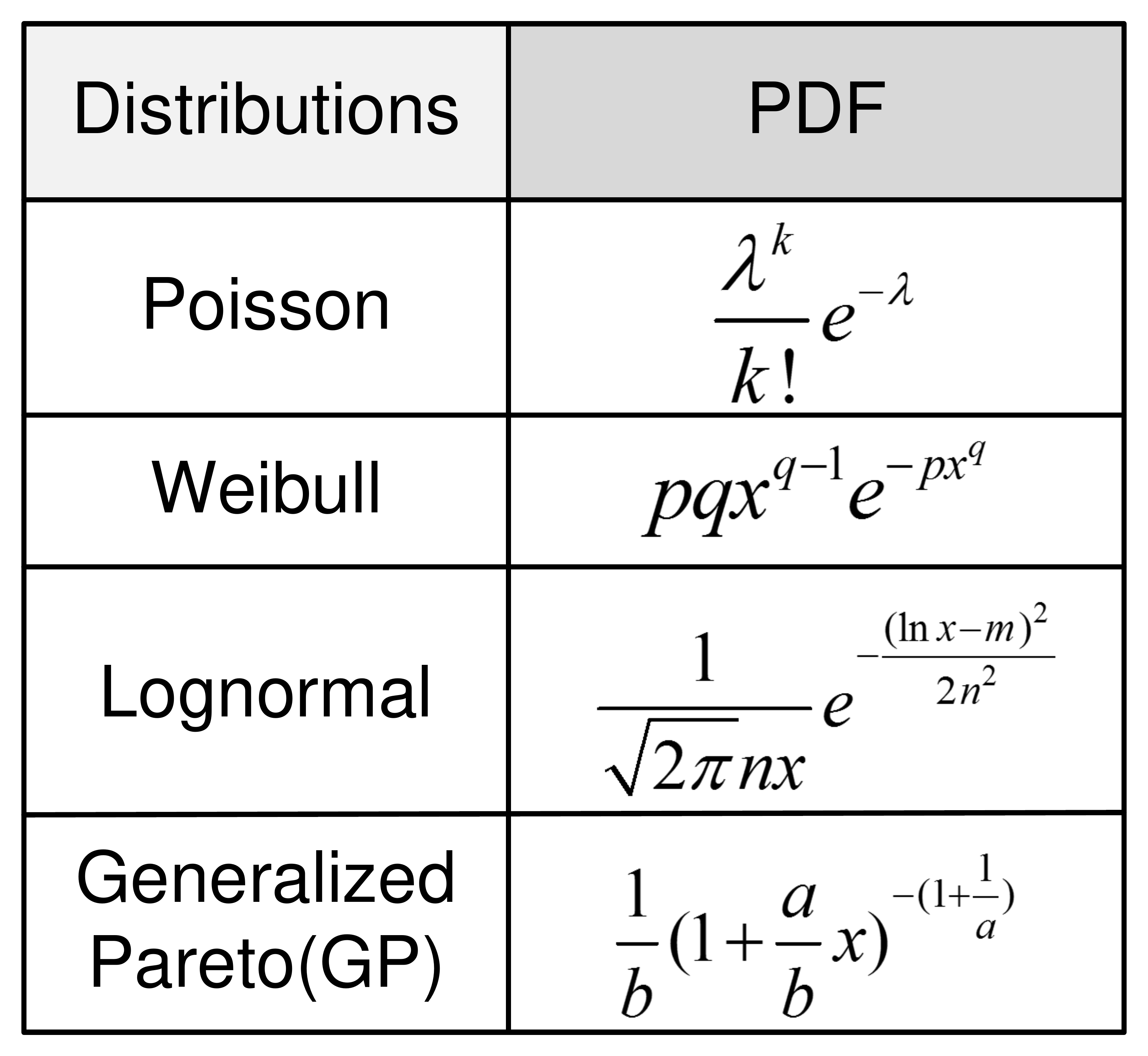}
	\caption{The candidate distributions and their PDF expressions.}
\end{table}

The real PDF and the fitted ones are presented in Fig. 4 for European cities and in Fig. 5 for Asian ones, respectively. In order to verify the best fitness for the Euler characteristics, the root mean square error (RMSE) between the real distribution and every candidate is listed in Table V, and it is obvious that the RMSE between log-normal distribution and the real PDF is almost one-order of magnitude smaller than the others in the same row. As a result, an astounding but well-grounded conclusion can be drawn as follows: regardless of the geographical boundaries, the culture differences and the historical limitations, the Euler characteristics of both Asian and European cities entirely conform to log-normal distribution.

\begin{figure}[ht]
	\centering
	
	\subfigure[London $ \qquad\qquad\qquad\qquad\qquad\qquad $(c) Paris]{
		\makebox[4cm][c]{
			\includegraphics[scale=0.13]{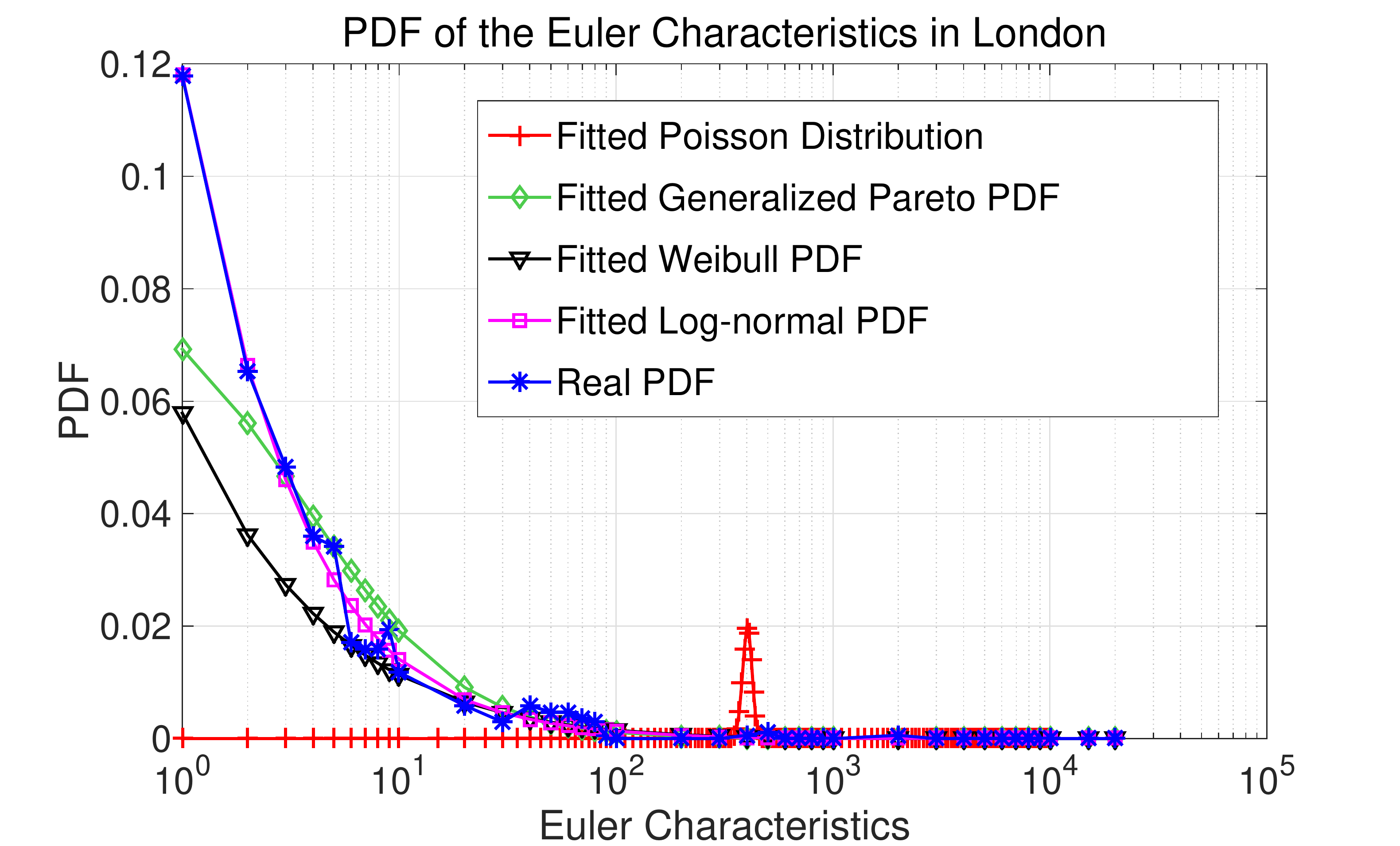}
		}
		
		\makebox[4cm][c]{
			\includegraphics[scale=0.13]{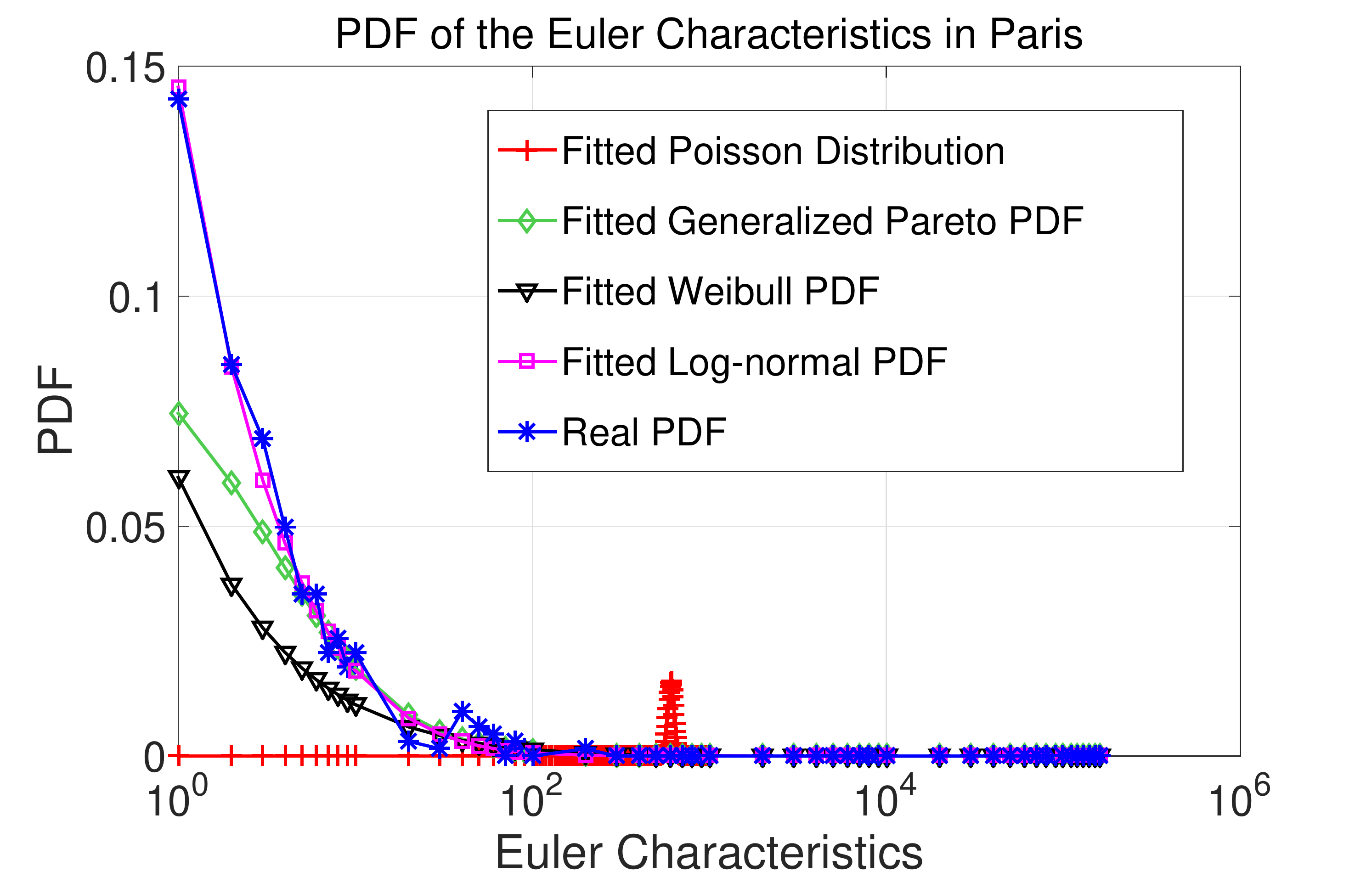}
		}
	}
	
	\subfigure[Munich $ \qquad\qquad\qquad\qquad\qquad\qquad $(d) Warsaw]{
		\makebox[4cm][c]{
			\includegraphics[scale=0.13]{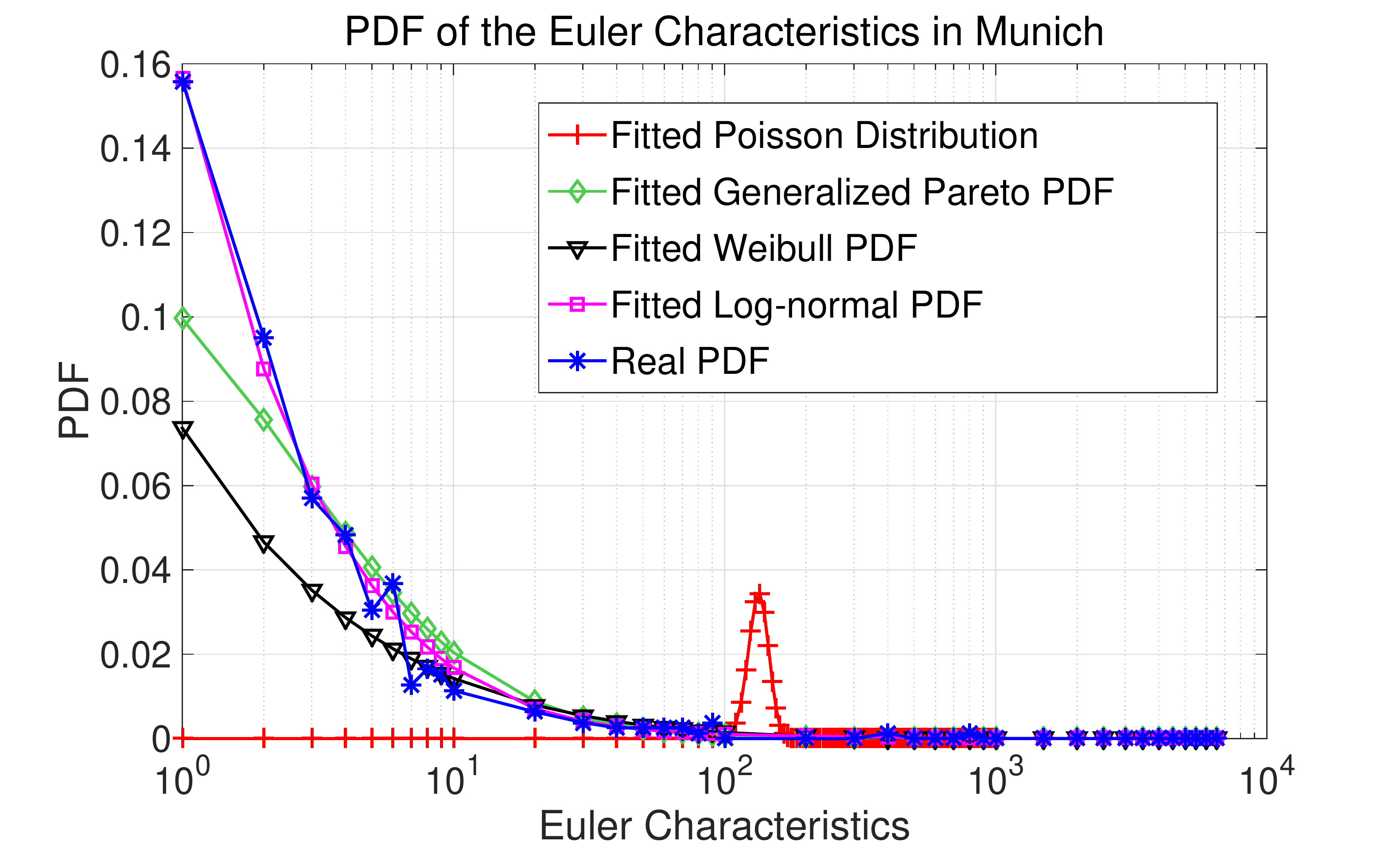}
		}
		
		\makebox[4cm][c]{
			\includegraphics[scale=0.13]{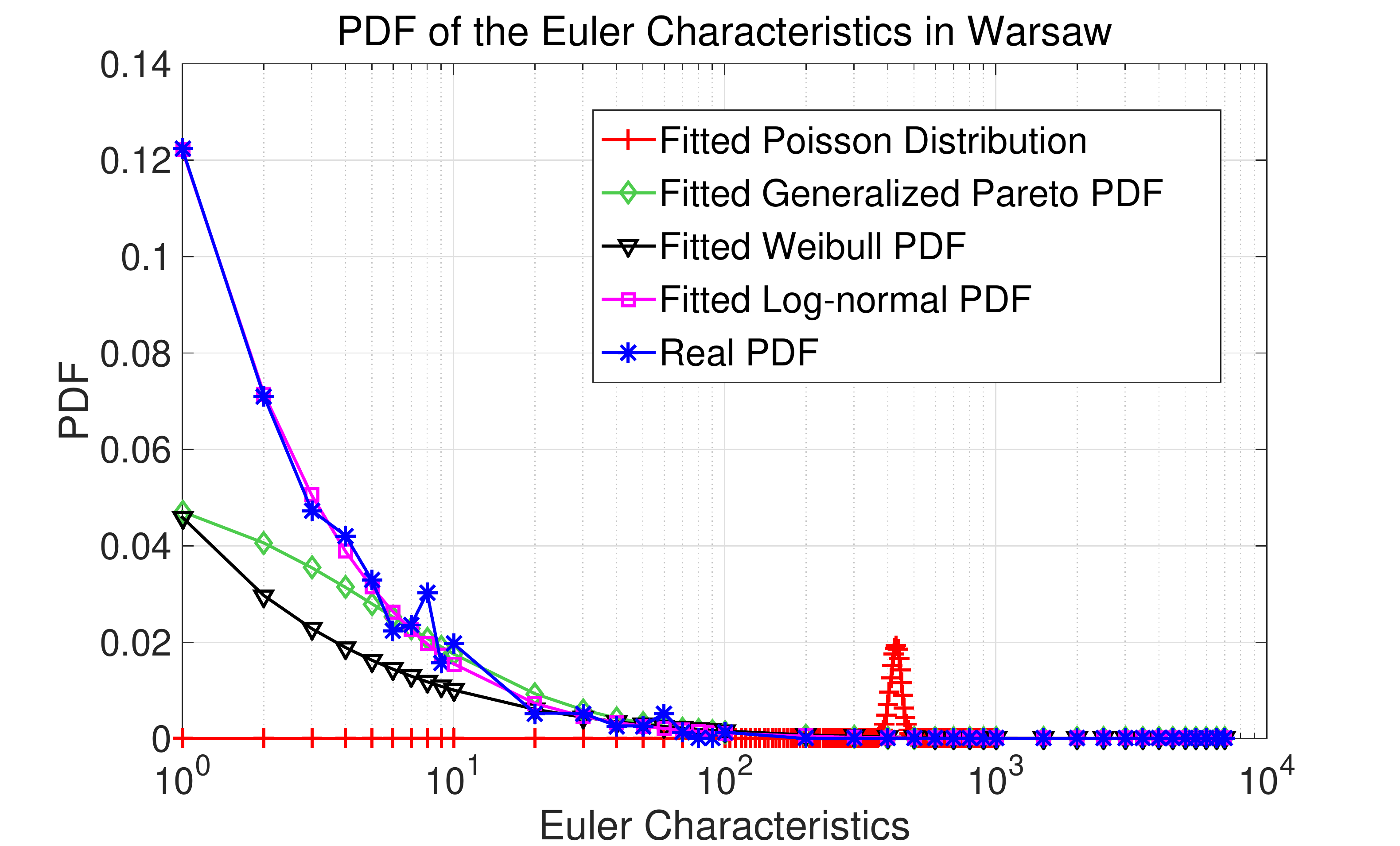}
		}
	}
	\caption{The comparison between the practical PDF and the fitted ones for the Euler characteristics of BS locations in European cities.}
\end{figure}

\begin{figure}[ht]
	\centering
	
	\subfigure[Beijing $ \qquad\qquad\qquad\qquad\qquad\quad $(c)Seoul]{
		\makebox[4cm][c]{
			\includegraphics[scale=0.13]{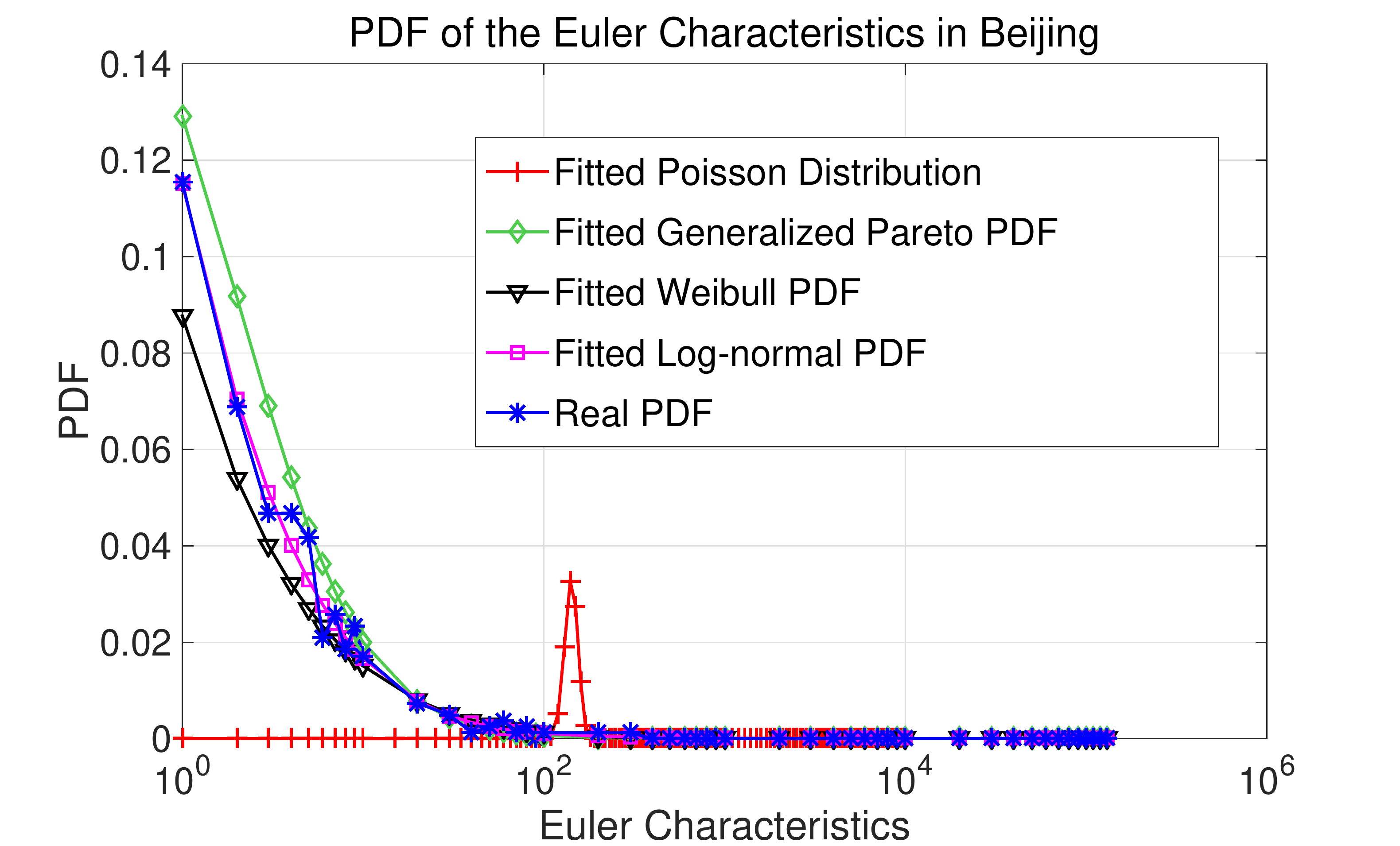}
		}
		
		\makebox[4cm][c]{
			\includegraphics[scale=0.13]{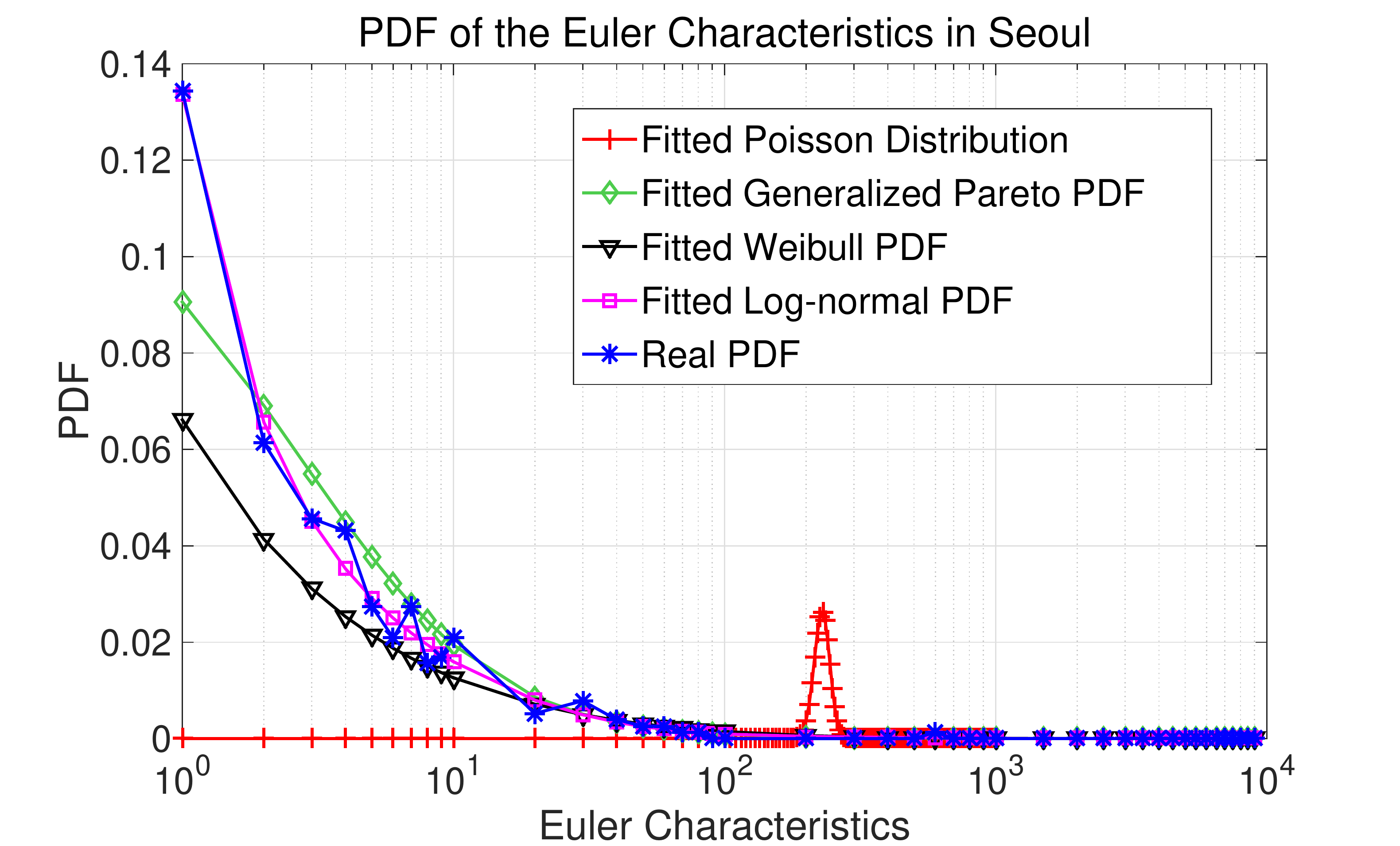}
		}
	}
	
	\subfigure[Mumbai $ \qquad\qquad\qquad\qquad\qquad\qquad $(d) Tokyo]{
		\makebox[4cm][c]{
			\includegraphics[scale=0.13]{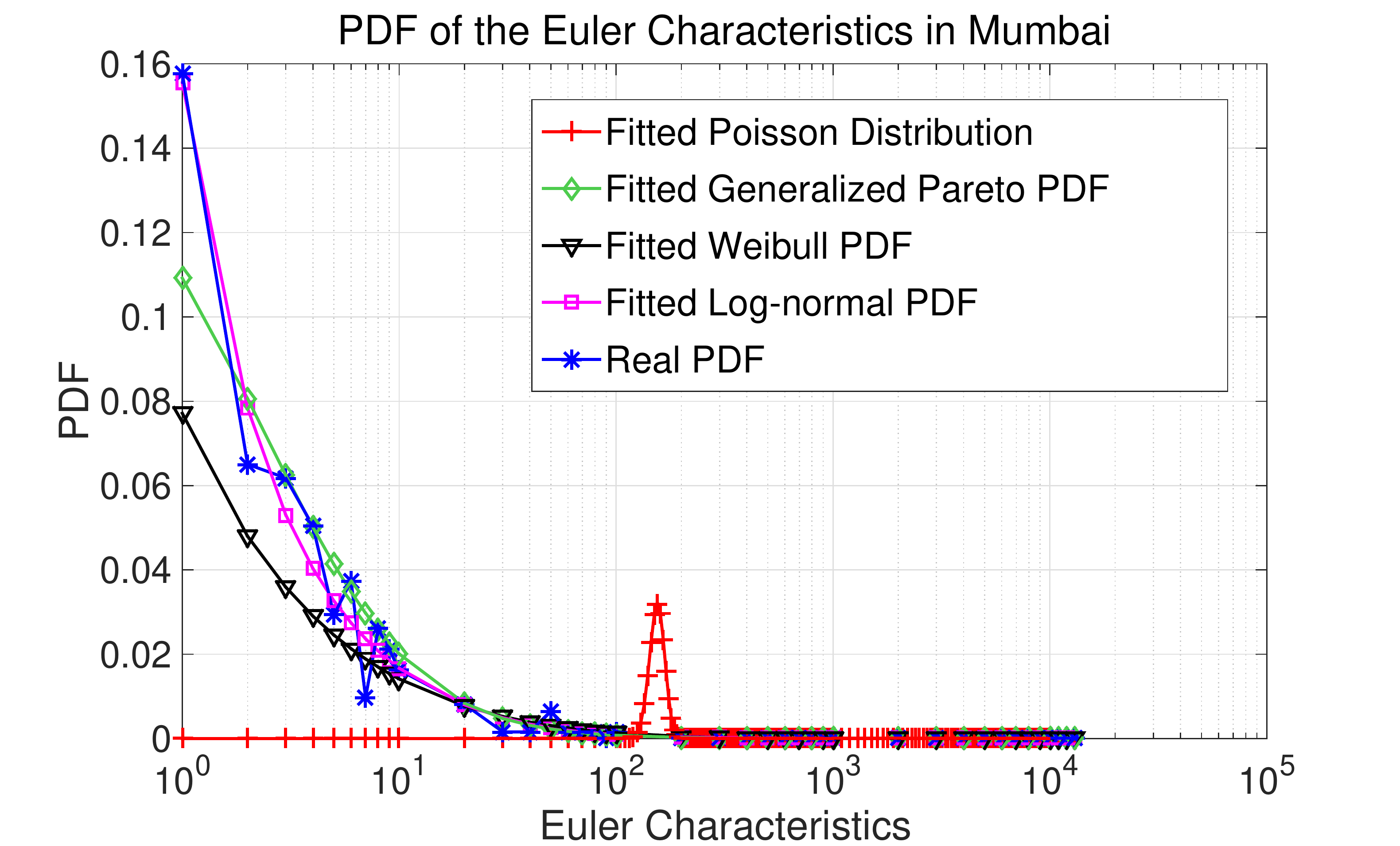}
		}
		
		\makebox[4cm][c]{
			\includegraphics[scale=0.13]{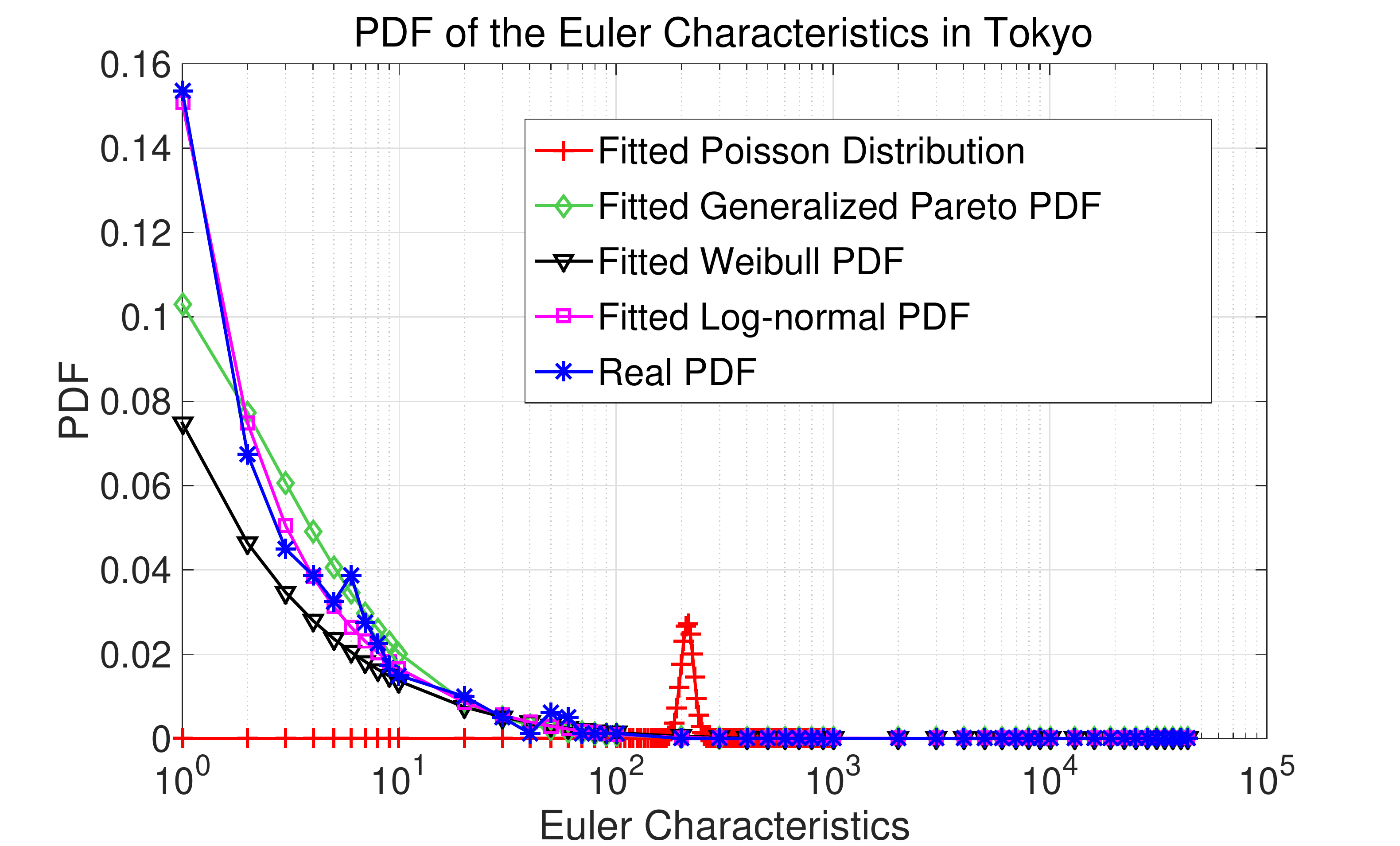}
		}
	}
	
	\caption{The comparison between the practical PDF and the fitted ones for the Euler characteristics of BS locations in Asian cities.}
\end{figure}

\begin{table}[h]
	\centering
	\includegraphics[scale=0.25]{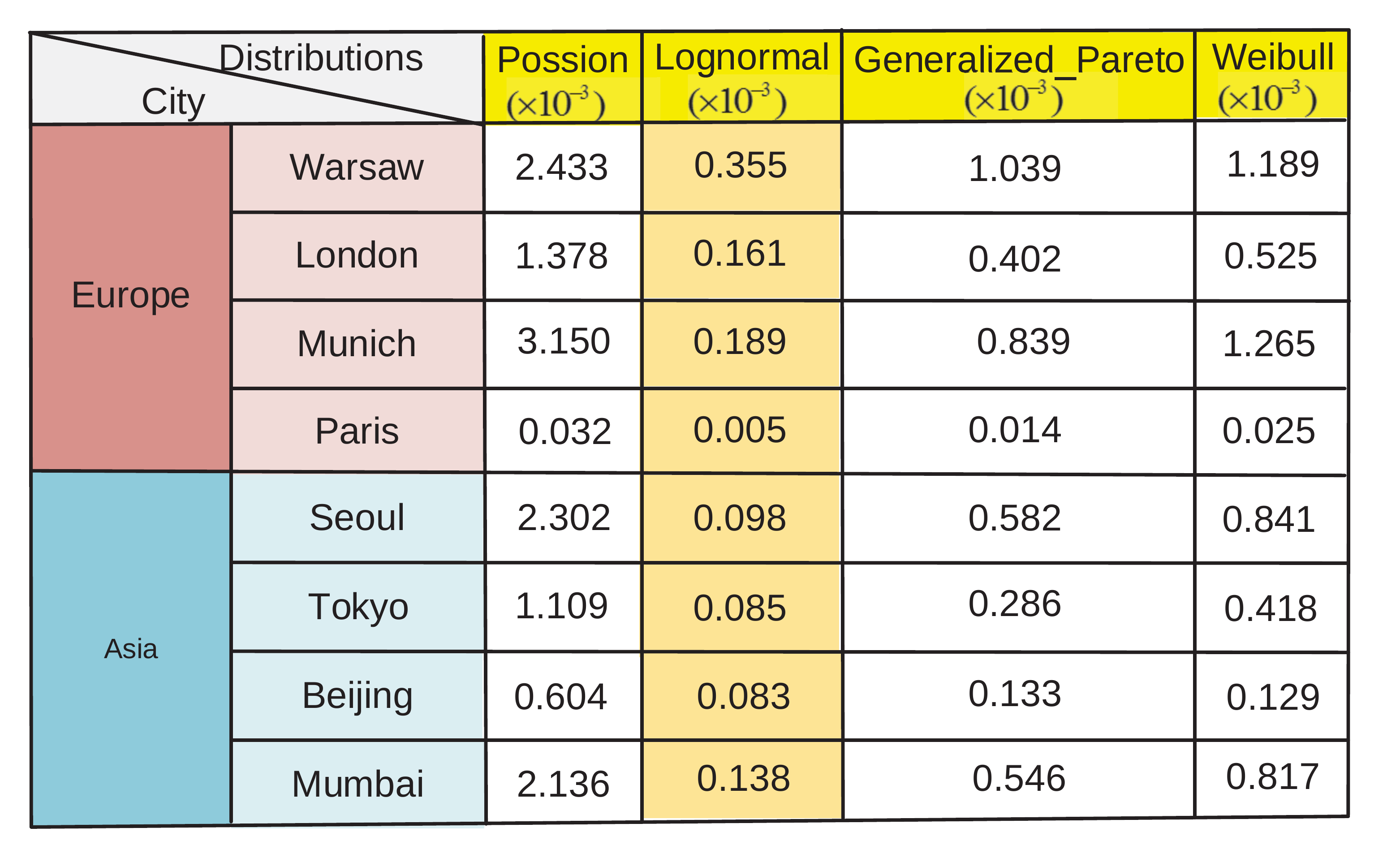}
	\caption{The RMSE between each candidate distribution and the practical one.}
\end{table}

\section{Conclusion and Future Works}
In this letter, several typical instruments in the algebraic geometric field, namely, $ \alpha $-Shapes, Betti numbers, and Euler characteristics, have been utilized in the discovery of inherent topological essences in BS deployments for Asian and European cities. Firstly, the fractal nature has been revealed in the BS topology according to both the Betti numbers and the Hurst coefficients. Moreover, it has been proved that log-normal distribution provides the best match for the PDFs of the Euler characteristics of the real BS location data in cellular networks among the classical candidate distributions.

However, regardless of the topological discoveries above, some thought-provoking problems still need to be solved. For example, how can the fractal nature be applied to the design of cellular networks? How can log-normal distribution of the Euler characteristics facilitate BS deployments? All of these issues will be investigated in our future works.

\bibliographystyle{IEEEtran}
\bibliography{IEEEfull,reference}

\end{document}